\documentclass[aps,prx,preprintnumbers,amsmath,amssymb,twocolumn,letterpaper,unsortedaddress,floatfix]{revtex4}
\usepackage{amssymb}
\usepackage{ifpdf}
\usepackage[utf8]{inputenc}
\usepackage{graphicx}
\usepackage{hyperref}
\usepackage{bm}
\usepackage{textcomp}
\usepackage{times}

\makeatletter
\newcommand*{\balancecolsandclearpage}{%
  \close@column@grid
  \clearpage
  \twocolumngrid
}
\makeatother

\usepackage{bibunits}

\providecommand{\vect}[1]{\boldsymbol{#1}}
\newcommand{\nn}{\nonumber}

\begin{document}
\title{Fluctuation-induced first-order phase transition in Dzyaloshinskii-Moriya helimagnets}

\author{M. Janoschek}
\affiliation{Physik Department E21, Technische Universit\"at M\"unchen, D-85748 Garching, Germany}
\affiliation{Department of Physics, University of California, San Diego, La Jolla, CA 92093-0354, USA}
\affiliation{Los Alamos National Laboratory, Los Alamos, New Mexico 87545, USA}
\email[Corresponding Author: ]{mjanoschek@lanl.gov}
\author{M. Garst}
\affiliation{Institute for Theoretical Physics, Universit\"at zu K\"oln, D-50937 K{\"o}ln, Germany}
\author{A. Bauer}
\affiliation{Physik Department E21, Technische Universit\"at M\"unchen, D-85748 Garching, Germany}
\author{P. Krautscheid}
\affiliation{Institute for Theoretical Physics, Universit\"at zu K\"oln, D-50937 K{\"o}ln, Germany}
\author{R. Georgii}
\affiliation{Forschungsneutronenquelle Heinz Maier-Leibnitz (FRM II), Technische Universit\"at M\"unchen, D-85748 Garching, Germany}
\affiliation{Physik Department E21, Technische Universit\"at M\"unchen, D-85748 Garching, Germany}
\author{P. B\"oni}
\author{C. Pfleiderer}
\affiliation{Physik Department E21, Technische Universit\"at M\"unchen, D-85748 Garching, Germany}

\date{\today}

\begin{abstract}
Two centuries of research on phase transitions have repeatedly highlighted the importance of critical fluctuations that abound in the vicinity of a critical point. They are at the origin of scaling laws obeyed by thermodynamic observables close to second-order phase transitions resulting in the concept of universality classes, that is of paramount importance for the study of organizational principles of matter. Strikingly, in case such soft fluctuations are too abundant they may alter the nature of the phase transition profoundly; the system might evade the critical state altogether by undergoing a discontinuous first-order transition into the ordered phase. Fluctuation-induced first-order transitions have been discussed broadly and are germane for superconductors, liquid crystals, or phase transitions in the early universe, but clear experimental confirmations remain scarce. Our results from neutron scattering and thermodynamics on the model Dzyaloshinskii-Moriya (DM) helimagnet (HM) MnSi show that such a fluctuation-induced first-order transition is realized between its paramagnetic and HM state with remarkable agreement between experiment and a theory put forward by Brazovskii. While our study clarifies the nature of the HM phase transition in MnSi that has puzzled scientists for several decades, more importantly, our conclusions entirely based on symmetry arguments are also relevant for other DM-HMs with only weak cubic magnetic anisotropies. This is in particular noteworthy in light of a wide range of recent discoveries that show that DM helimagnetism is at the heart of problems such as topological magnetic order, multiferroics, and spintronics.
\end{abstract}

\vskip2pc

\maketitle

\begin{bibunit}
\section{Introduction}

Critical phenomena were observed for the first time in 1822 in the form of the critical opalescence of water vapor when Cagniard de la Tour discovered the critical point of the gas to liquid phase transition \cite{Tour:1822}. The associated change of the physical properties is also referred to as continuous (or second-order) transition, implying that the order parameter characterizing the ordered state, emerges smoothly. Hence, as recognized by Cagniard de la Tour a critical point represents a ``special state'' (\textit{\'{e}tat particulier}) \cite{Tour:1823} because the disordered and the ordered phase are indistinguishable.

The unusual physical properties are thereby generally referred to as \textit{critical phenomena}. At the heart of the critical phenomena is an abundance of low-energy fluctuations of the order parameter that extend over increasing length scales as the critical point is approached. The divergence of their so-called correlation length $\xi$ results in universal scaling laws for observables that only depend on the symmetries of the critical system while being independent of its specific microscopic details. This led to the notion of \textit{universality classes} --- a cornerstone of modern physics --- providing a common framework for a wide range of systems with the same critical behavior despite entirely different microscopic character \cite{Wilson:83}.

As one of the most remarkable aspects, it has long been noticed theoretically that an excess of critical fluctuations may change the nature of the phase transition entirely. If the phase space available for the critical degrees of freedom is sufficiently large, the system may evade the critical point to avoid the large entropy associated with the fluctuations by realizing a discontinuous first-order transition into the ordered phase. As a result, the correlation length does not diverge and the order parameter varies also discontinuously at the transition which is then accompanied by the release of latent heat.

Theoretically, fluctuation-induced first-order transitions occur as a consequence of non-analytic terms in the Ginzburg-Landau free energy functional that are generated by fluctuation-induced corrections. Perhaps the best-known example concerns thereby the Coleman-Weinberg effective potential for an order parameter coupled to a fluctuating gauge field \cite{Coleman:73}. This found important applications in the context of phase transitions in the early universe addressing cosmic inflation \cite{Linde:82} and the problem of baryogenesis \cite{Trodden:99}. However, it is also relevant for the description of superconductors and smectic-A liquid crystals \cite{Halperin:74}. In contrast, for an order parameter with a large number of components $N \geq 4$, its self-interaction may already suffice to drive the transition first-order as pointed out by Bak and coworkers \cite{Bak:76}. Finally, Brazovskii \cite{Brazovski:75} considered theoretically critical fluctuations that become soft not only at a single point in momentum space (as e.g. for a ferromagnet), but rather on a finite manifold, notably a sphere. In such a case, the density of states for critical fluctuations exhibits a one-dimensional singularity so that interaction corrections are expected to drive a strong suppression of the correlation length and, eventually, a fluctuation-induced first-order transition.

Various settings have been proposed in which Brazovskii-type transitions may be expected such as weakly anisotropic antiferromagnets \cite{Brazovski:75}, liquid crystals \cite{Brazovski:76,Swift:76}, diblock copolymers \cite{Leibler:80}, the Rayleigh-B\'{e}nard convective instability \cite{Swift:77}, pion condensation in nuclear matter \cite{Migdal:90}, and Bose-Einstein condensates in multimode cavities \cite{Gopalakrishnan:09} or with spin-orbit coupling \cite{Gopalakrishnan:11}. However, only few experimental realizations of Brazovskii systems such as diblock copolymers \cite{Bates:90} were reported so far. The outstanding experimental constraint thereby are precision measurements of the relevant fluctuations, which are difficult to resolve \cite{Binder:87}. Prior to the present study direct experimental evidence for a Brazovskii-transition hence represented a major challenge of interest to a wide range of topics.

In this paper we identify the onset of helimagnetic (HM) order in a class of cubic materials as a prime example for a Brazovskii transition using a quantitative comparison of experiment and theory. Specifically we consider systems, in which the HM order arises from (chiral) Dzyaloshinsky-Moriya (DM) spin-orbit interactions in the presence of weak cubic magnetic anisotropies. Prominent members of this class of DM-HMs crystallize in the non-centrosymmetric P2$_1$3 space group, encompassing the metallic B20 compounds MnSi and FeGe, the B20 semiconductor Fe$_{1-x}$Co$_x$Si, and the multiferroic insulator Cu$_2$OSeO$_3$. These materials have recently generated tremendous scientific interest, when a new form of magnetic order -- a skyrmion lattice -- was discovered \cite{Muehlbauer:09, Muenzer:10, Yu:10,Yu:11}.  As its novel aspect each skyrmion is characterized by a non-zero topological winding number. Moreover, in the skyrmion phase of MnSi spin transfer torques have been observed at record low current densities $j = 10^6$~A m$^{-2}$ making these materials interesting for spintronic applications \cite{Jonietz:10, Schulz:12}. In the insulator Cu$_2$OSeO$_3$, magnetic skyrmions  possess electric polarization and the concomitant magnetoelectric coupling promises interesting multiferroic behavior \cite{Seki:12,Adams:12}.

In fact, it was argued in Ref.~\cite{Muehlbauer:09} that the skyrmion lattice phase only exists due to strong renormalizations attributed to thermal fluctuations. This motivates a more detailed study of fluctuation effects on the phase diagram in general. In the present work, we investigated the influence of fluctuations on the HM transition of MnSi at zero field by means of carefully designed experiments using a combination of small angle neutron scattering (SANS), and measurements of the magnetic susceptibility and specific heat. Notably, we have mapped out the fluctuations in three dimensions demonstrating that they indeed emerge on a Brazovskii-sphere. Within Brazovskii theory the temperature dependence of the correlation length hereby naturally explains all features observed in the thermodynamic measurements. As our results are entirely based on symmetry considerations they are expected to be of general importance for DM-HMs.

Our paper is organized as follows; first, we review the state-of-the-art on the HM phase transition in MnSi and the experimental methods used for this work, followed by a description of our experimental and theoretical results, before we conclude with a discussion.

\begin{figure*}[th!]
\centering
\includegraphics[width=.8\textwidth]{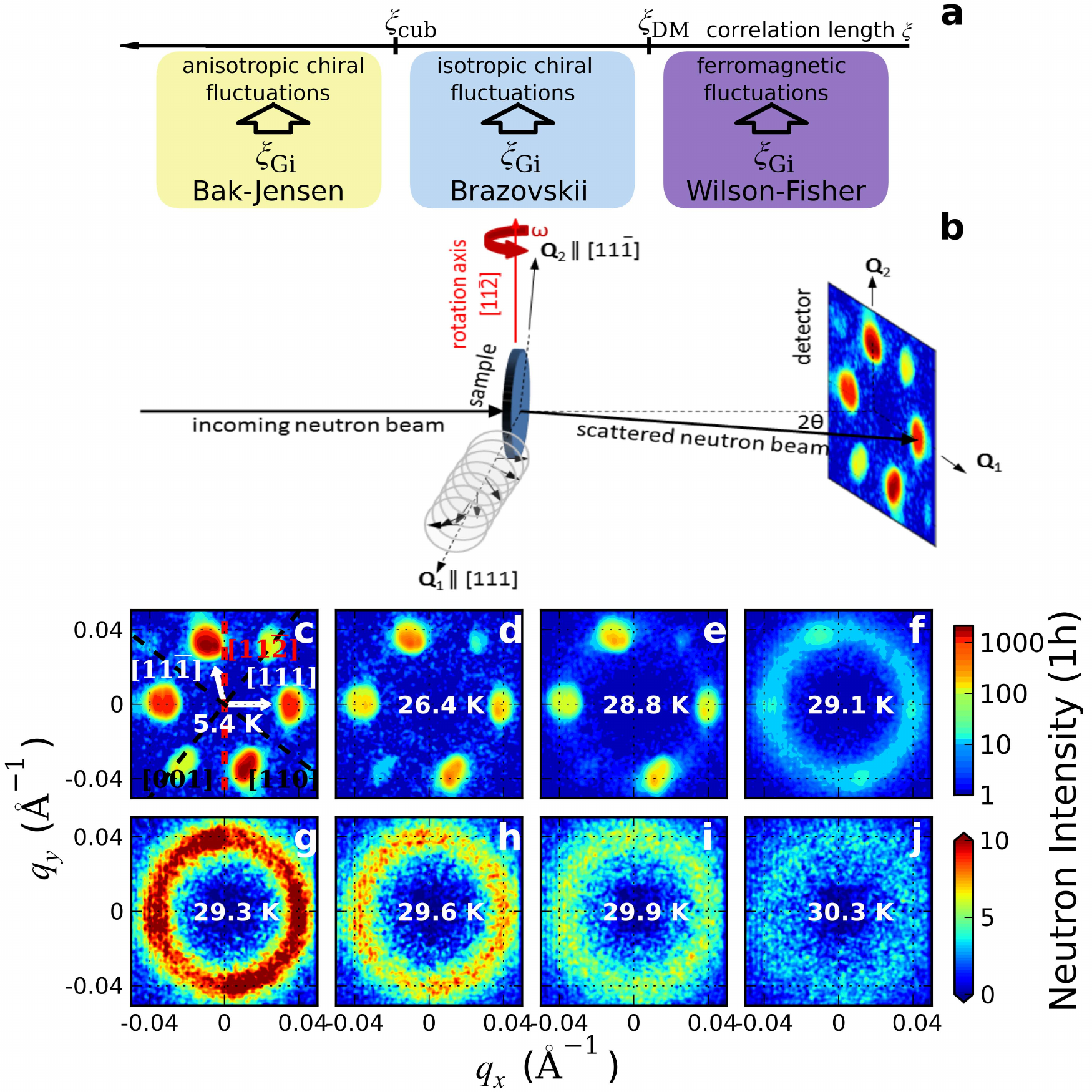}
\caption{Helimagnetic (HM) phase transition of MnSi at $T_c =29$K investigated by small angle neutron scattering (SANS). (a) There are two crossovers for non-interacting fluctuations in a Dzyaloshinskii-Moriya (DM) HM as $T_c$ is approached and the correlation length $\xi$ increases. Interactions result in an additional Ginzburg length $\xi_{\rm Gi}$ and depending on its size different scenarios are realized (see text). (b) The SANS setup used to study the temperature-dependence of $\xi$, where the neutron beam scatters off the magnetic helix. Turning the sample around the rotation axis parallel to the $[11\overline{2}]$ zone axis (red arrow) the sample can be rocked through the Bragg condition for the magnetic helix. With this sample orientation the magnetic Bragg condition can be fulfilled for the two magnetic propagation vectors $\mathbf{Q}_1$ = $[111]$ and $\mathbf{Q}_2$ = $[11\overline{1}]$ [white arrows in panel (c)] corresponding to two of the four possible helical $Q$-domains (see text). (c)-(j)
Magnetic intensity distribution of MnSi close to $T_c$. Below $T_c$ [panels (c)-(e)] discrete magnetic Bragg spots corresponding to the helical order are visible, whereas above $T_c$ [panels (g)-(j)] the magnetic intensity spreads out over a sphere in reciprocal space. The background scattering was determined well above the magnetic phase transition and was subtracted from all data sets. The black broken lines in (c) indicate weak reflections due to multiple scattering effects along the [001] and [110] directions. Note, that the intensity in panels (c)-(f) is plotted on a logarithmic scale for a better comparison between the intensities on the helical satellites and the sphere.}\label{fig:Fig1}
\end{figure*}

\section{The Helimagnetic Phase Transition in MnSi}

In the absence of a magnetic field MnSi exhibits a HM ground state below $T_c \approx 29$ K \cite{Bak:80,Ishida:85} characterized by a hierarchy of energy scales due to the weak spin-orbit coupling $\lambda_{\rm SO}$. First, the strong ferromagnetic exchange interaction $J$ aligns the spins on short length scales. Second, the lack of inversion symmetry in the $P 2_1 3$ space group allows for a weak DM interaction, $D \sim \mathcal{O}(\lambda_{\rm SO})$, that leads to a chiral twist of the magnetization and stabilizes HM order with a pitch of 180~\AA\ deep in the ordered phase. Third, weak cubic anisotropies that arise in higher order in $\lambda_{\rm SO}$ finally lock the magnetic helix in a cubic $\langle 111 \rangle$ direction~\cite{Bak:80}. Importantly, this hierarchy of energy/length scales is also reflected in the nature of critical fluctuations as $T_c$ is approached. For $T \gg T_c$ the correlation length $\xi$ is short and the fluctuations have an essential ferromagnetic character. However, as the temperature is lowered and $\xi$ reaches the order of the DM length scale $\xi_{\rm DM} \sim J/D$, the fluctuations start to accumulate uniformly on a sphere in momentum space of radius $Q = D/J$ and, as a consequence, the magnetic correlations develop an oscillating character. Finally, as the correlation length increases even further, $\xi \gtrsim \xi_{\rm cub}$, cubic anisotropies favor the fluctuations to carry momentum in the crystallographic $\langle 111 \rangle$ directions.

While the HM transition is expected to be second-order on a mean-field level, interactions between the HM fluctuations were theoretically predicted to give rise to important corrections driving the transition first-order. The precise mechanism, however, depends crucially on the strength of the interaction that generates an additional scale, i.e., the Ginzburg length $\xi_{\rm Gi}$ \cite{ChaikinBook}, see Fig.~\ref{fig:Fig1}(a). According to the Ginzburg criterion, interactions can be treated perturbatively in the case of short correlations, $\xi \ll \xi_{\rm Gi}$, while the fluctuations are strongly interacting if $\xi \gtrsim \xi_{\rm Gi}$. The limit of very weak interactions, $\xi_{\rm Gi} \gg \xi_{\rm cub} \gg \xi_{\rm DM}$, was considered by Bak and Jensen \cite{Bak:80} who argued that in the regime $\xi \gg \xi_{\rm cub}$ an effective field-theoretical description emerges in terms of an order parameter with $N=8$ components corresponding to amplitude and phase of helices with momenta along the four equivalent $\langle 111 \rangle$ directions. The residual interaction between order parameter fluctuations might then drive the transition first-order. On the other hand, if the cubic anisotropies are associated with the smallest energy scale, i.e., the largest length scale, $\xi_{\rm cub} \gg \xi_{\rm Gi} \gg \xi_{\rm DM}$, HM fluctuations are already strongly interacting while they are still uniformly distributed on a sphere in momentum space. This scenario coincides with the situation considered by Brazovskii \cite{Brazovski:75}, and an interaction-induced first-order transition preempts the cubic crossover in this case. For the zero-temperature HM transition that is observed in MnSi as function of pressure, the corresponding Brazovskii scenario was studied theoretically by Schmalian and Turlakov \cite{Schmalian:04}. Finally, a third scenario arises for very strong interactions, $\xi_{\rm cub} \gg \xi_{\rm DM} \gg \xi_{\rm Gi}$, where the fluctuations are already strongly interacting before they develop a preferred chirality. Here, the physics on length scales $\xi_{\rm DM} > \xi > \xi_{\rm Gi}$ is governed by the Wilson-Fisher renormalization group fixed-point resulting in a universal DM crossover at $\xi \sim \xi_{\rm DM}$. We demonstrate below that for MnSi the Brazovskii scenario is realized.

Although MnSi has been the topic of intense scientific study for decades, the nature of the transition from the paramagnetic (PM) to the HM phase at zero field is still debated controversially. Initially, the transition was interpreted to be of second-order based on the specific heat in polycrystals \cite{fawcett:1970}, an abundance of paramagnon fluctuations also observed in neutron scattering \cite{Ishikawa:85,Roessli:02,Grigoriev:05}, as well as a pronounced Curie-Weiss behavior in the magnetic susceptibility up to $T_c$ \cite{Pfleiderer:97}. However, a careful analysis of the temperature dependence of the HM Bragg peaks suggested that the transition has a weak first-order character \cite{Bernhoft:94}. The discovery of partial magnetic order in MnSi at high pressures~\cite{Pfleiderer:04} inspired a search for spin textures with non-trivial topology. In particular, the observation of a broad shoulder at $T^*$~$\approx$~$T_c+1$~K, revealed in more detailed measurements of the specific heat~\cite{Pfleiderer:JMMM2001}, led to the theoretical proposal of a skyrmion liquid phase as a generic precursor phenomenon at essentially each PM to HM transition in DM-HMs \cite{roessler:06}. This interpretation was adopted by Pappas \textit{et al.} \cite{Pappas:09,Pappas:11} who argued that the so-called chiral fraction determined via polarized neutron scattering studies provides evidence for a skyrmion-liquid phase. Similarly, Hamann {\it et al.} qualitatively explained their neutron scattering results as a complex form of long-range order, a so-called magnetic blue phase, motivated by an amorphous lattice of skyrmions observed theoretically \cite{Hamann:11}. However, on closer inspection all claims of a skyrmion liquid are highly debatable since they either involve unconventional terms in the Landau theory, do not account for the relationship of chiral fraction with skyrmion liquid or, last but not least, were obtained with Monte Carlo calculations on relatively small system sizes, respectively.

As an alternative, several studies suggested that the unusual specific heat anomaly in MnSi can be explained in the traditional framework of HM fluctuations in the presence of cubic anisotropies. For instance, in 2007 Stishov {\it et al.} \cite{Stishov:07} demonstrated by means of the specific heat, thermal expansion and electrical resistivity that the HM transition is indeed first-order with a tiny latent heat. These authors emphasized that the broad shoulder in the specific heat, noticed earlier \cite{Pfleiderer:JMMM2001,roessler:06}, might be caused by HM fluctuations. Yet, the origin of the first-order transition remained unresolved. Further, Bauer {\it et al.}~\cite{Bauer:10} showed that the specific heat close to $T^*$  exhibits a so-called Vollhardt invariance \cite{Vollhardt:97}, i.e., a crossing point that is invariant under small magnetic fields. Finally, Grigoriev {\it et al.}~\cite{Grigoriev:11} interpreted a corresponding signature at $T^*$ in the magnetic susceptibility and the correlation length as a crossover in the character of the fluctuations as the HM transition is approached.

\begin{figure}[th!]
\includegraphics[width=.5\textwidth]{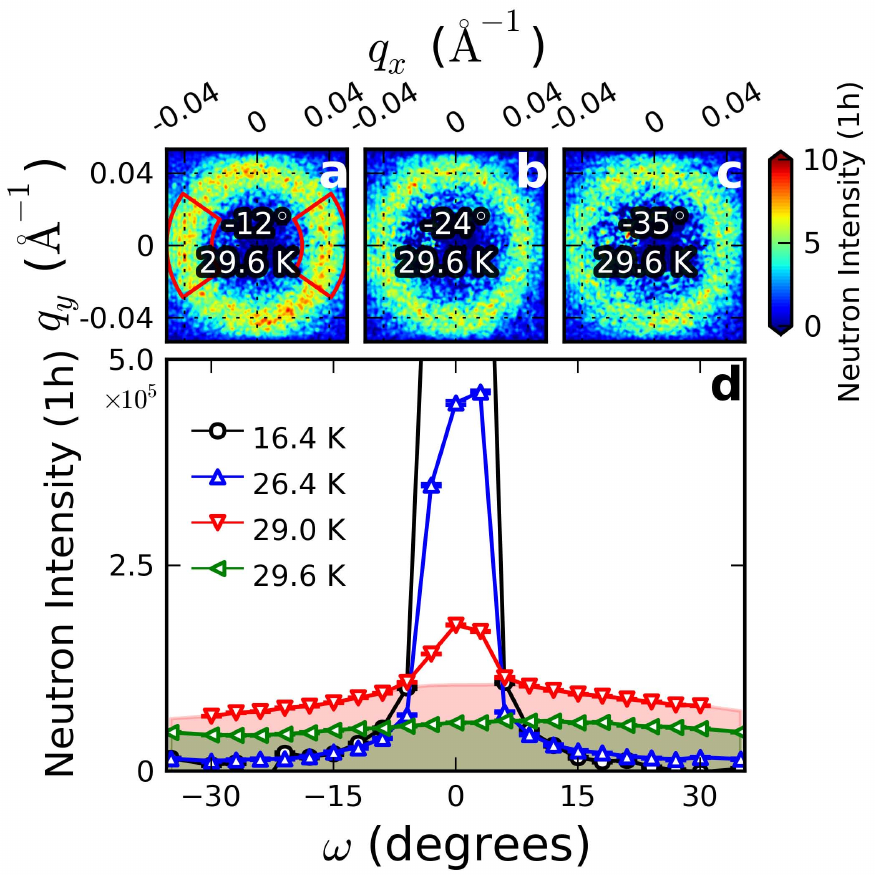}
\caption{(a)-(c) Dependence of the magnetic fluctuations on the rocking angle $\omega$ up to $\pm$35$^\circ$ are exemplaryily shown for $T$~=~29.6~K. The sample was rocked around the  $[11\overline{2}]$-axis [see Fig.~\ref{fig:Fig1}(a)]. (d) Integrated intensities corresponding to the red integration regions in panel (a) as a function of $\omega$ for different temperatures above and below $T_c$. The integration regions were selected such that all included points are rotated through the detector plane with approximately identical velocity. Further, the integrated intensities have been corrected for the change of neutron absorption due to the increasing path length through the sample for increasing angles. The red and green shaded regions denote the integrated intensity for the critical magnetic fluctuation at 29~and 29.6~K, respectively.}\label{fig:Fig2}
\end{figure}

\begin{figure*}[th!]
\includegraphics[width=\textwidth]{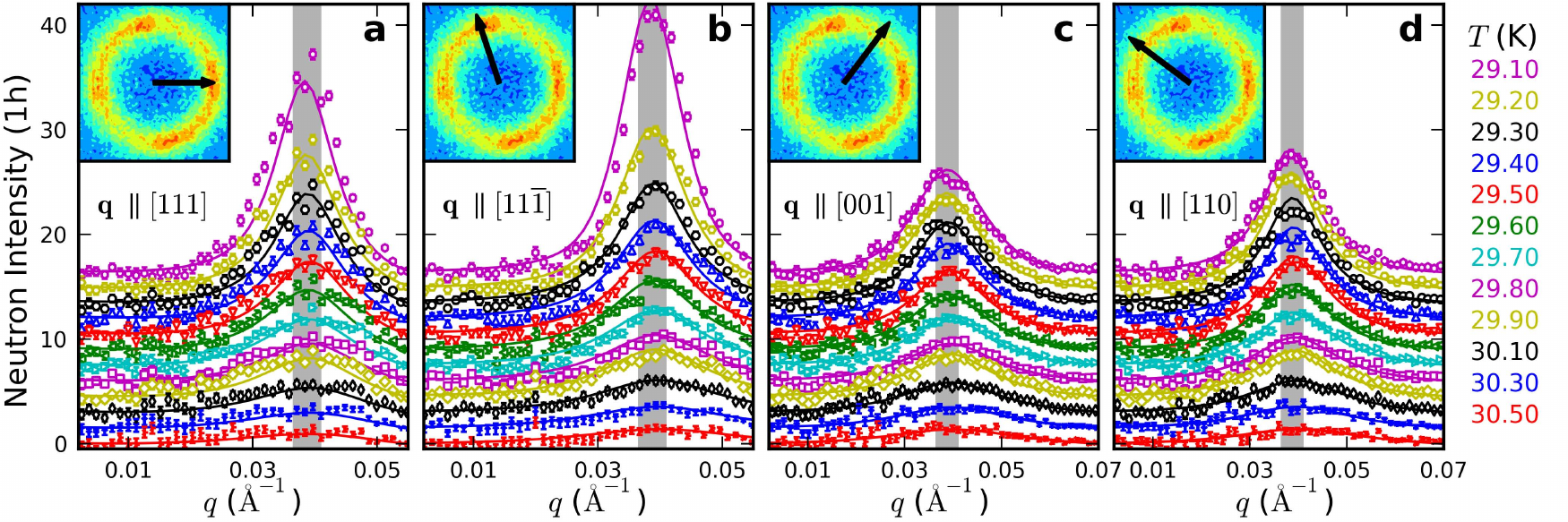}
\caption{Radial $\mathbf{q}$-scans through the magnetic fluctuations above $T_c$~$=$~29 K are shown for all measured temperatures. The inset in each panel shows the corresponding direction of the scan in reciprocal space. The $\mathbf{q}$-scans were extracted from the two-dimensional detector images in Fig.~\ref{fig:Fig1} by performing radial bins with an azimuthal width corresponding to the experimental resolution. The solid lines are fits to Eq.~\eqref{eq:mnsi_ring_scattering} (see text). The grey shaded region shown in each panel denotes the experimental resolution.}\label{fig:Fig3}
\end{figure*}

\section{Experimental Methods}

The small angle neutron scattering (SANS) experiment was performed using neutrons with wavelength $\lambda$ = 9.7~\AA~ on the beamline MIRA \cite{Georgii:07} situated at the Forschungsneutronenquelle Heinz Maier-Leibnitz (FRM II). The instrumental resolution was selected by a computer-controlled variable source aperture of rectangular cross-section that was installed after the monochromator and a cadmium aperture of approximately 15~mm diameter in front of the sample and is comparable with that of other studies~\cite{Muehlbauer:09, Pappas:11}. The direct beam was masked by a Cd mask in front of the detector. For the experiment the single crystal was cooled with a closed cycle cryostat with sample tube (CCR) available at FRM II. The single crystal used for the SANS measurement is a disc of diameter $\approx$~20 mm and 2 mm thickness and was cut from a ingot that had already been studied before in several neutron experiments and was grown by the Bridgeman technique. \cite{Roessli:02,Janoschek:10}. The residual resistivity ratio (RRR) of this sample is $\approx$~80.

The specific heat and ac susceptibility were measured with a Quantum Design Physical Property Measurement System at temperatures down to 2 K and in magnetic fields up to 9 T. The specific heat was measured with a standard heat-pulse method, where heat pulses were limited to $\approx$ 0.2 \% to prevent thermal smearing of the signature of the first order transition. For both measurements the magnetic field $H$ was applied  along a $\langle$110$\rangle$ axis. The single crystal employed for the bulk measurements was grown by optical float zoning as detailed in Ref.~\cite{Bauer:10} and was cut into a cuboid shape (2$\times$2$\times$1.5 mm$^3$) with the two large surfaces aligned perpendicular to a $\langle$110$\rangle$ axis. The corresponding mass of this sample is 36.75 mg and the RRR is 80.

\begin{figure*}[th!]
\centering
\includegraphics[width=0.9\textwidth]{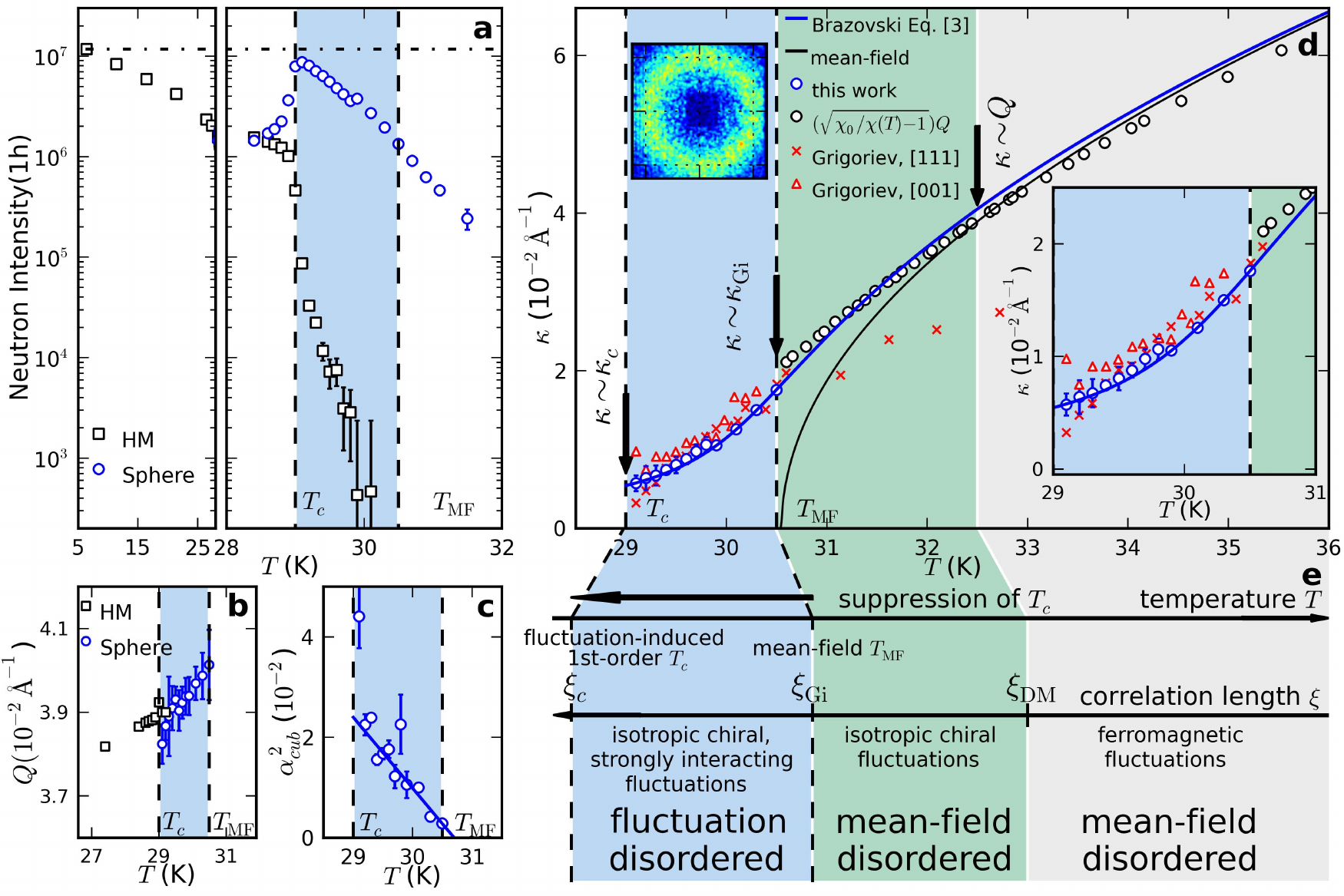}
\caption{Characteristics of the Brazovskii-type fluctuation-induced first-order transition of MnSi determined by SANS.  Parameters in panels (b)-(d) were obtained from the magnetic intensity by means of Eq.~\eqref{eq:mnsi_ring_scattering} (cf. Fig.~\ref{fig:Fig3}). (a) The magnetic intensity on the helical Bragg reflections (black squares) exhibits a discontinuous jump at $T_c$ as expected for a first-order transition. Above $T_c$ the magnetic intensity of the helical state is recovered entirely in form of critical magnetic fluctuations arising on a sphere in momentum space (blue circles).
(b) Magnitude of the helical propagation vector $Q$ in the HM phase (black squares) and the radius of the sphere above $T_c$ (blue circles) vs. temperature $T$. (c) Cubic anisotropy $\alpha^2_{\rm cub}$ vs. $T$.  (d) The inverse magnetic correlation length $\kappa= 2\pi/\xi$ vs. $T$ (blue circles). The red markers denote $\kappa$ according to previously published results \cite{Grigoriev:05}. The blue solid line is a fit to Eq.~\eqref{eq:kappa_bra} that describes the renormalization of $\kappa(T)$ for $T \rightarrow T_c$ as expected for a Brazovskii transition (see text). The black solid line highlights that for $T\gg T_{\rm MF}$ mean-field behavior is obtained. The black circles are obtained from measurements of the magnetic susceptibility via Eq.~\eqref{eq:chi} as described in the text. (e) The $T-$dependence of $\kappa$ identifies three separate regimes above $T_c$; (i) for $\xi < \xi_{\rm DM}$ [$\kappa > Q$, see (c)] fluctuations are essentially ferromagnetic as previously shown in Refs.~\cite{Ishikawa:85, Roessli:02};  (ii) they develop an isotropic chiral character for $\xi \gtrsim \xi_{\rm DM}$ as shown in the inset of (d);  (iii) for  $\xi > \xi_{\rm Gi}$ [$\kappa < \kappa_{\rm Gi}$] the interaction suppresses the transition temperature resulting in a fluctuation-disordered regime just above the fluctuation-induced first-order transition at $T_c$. We find for MnSi $T_c \approx 29$~K and $T_{\rm MF} \approx 30.5$~K and thus a suppression of $\Delta T = T_{\rm MF} - T_c \approx 1.5$~K.}\label{fig:Fig4}
\end{figure*}

\section{SANS close to the helimagnet transition of MnSi}

We start with a qualitative discussion of our SANS measurements before turning to a quantitative analysis.
Within the ordered phase, the helices carry momentum ${\bf Q}$ parallel to one of the four cubic $\langle 111 \rangle$ directions resulting in the formation of four domains. In our experiment [Fig.~\ref{fig:Fig1}(b)] the sample was oriented such that the Bragg condition is fulfilled for the two pairs of HM Bragg satellites associated with the domains $\mathbf{Q}_1 \parallel [111] $ and $\mathbf{Q}_2 \parallel [11\bar{1}]$ [white arrows in Fig.~\ref{fig:Fig1}(c)]. Below $T_c$ only the discrete Bragg satellites are visible [Fig.~\ref{fig:Fig1}(c)-(e)], but when the temperature approaches $T_c$ [Fig.~\ref{fig:Fig1}(e)] the magnetic intensity drops significantly and the Bragg spots start to broaden azimuthally. For $T$~$\geq$~$T_c$ magnetic intensity emerges on a ring with a radius that corresponds to the modulus, $Q$, of the helical propagation vector [Fig.~\ref{fig:Fig1}(f)-(j)].

To verify the Brazovskii scenario, we have determined the magnetic intensity distribution in three dimensions of momentum space via rocking scans that were carried out by rotating the sample around its vertical axis [Fig.~\ref{fig:Fig1}(b)]. In Fig.~\ref{fig:Fig2}(a)-(c), we show representative detector images for rocking angles $\omega$ up to 35$^\circ$ for $T$~$=$29.6~K. The magnetic intensities for all measured rocking angles are similar, showing that the fluctuations indeed emerge on a sphere. The magnetic anisotropy on the sphere is investigated in Fig.~\ref{fig:Fig2}(d) which shows integrated intensities as function of $\omega$ for several temperatures below and above $T_c$. The intensity increases significantly at large $\omega$ as soon as the critical magnetic fluctuations arise at $T_{c}$ [Fig.~\ref{fig:Fig2}(d), red triangles]. Here the sharp peak at the center is due to remanent HM order (red triangles, see discussion below), but also the intensity of the fluctuating part [Fig.~\ref{fig:Fig2}(d), red shaded region] initially exhibits a shallow maxima at the position of $\mathbf{Q}_1$ ($\omega$~$=$~0) due the cubic magnetic anisotropy. However, as $T$ is further increased the distribution of magnetic intensity on the sphere becomes more and more isotropic [Fig.~\ref{fig:Fig2}(d), green shaded region].

In Fig.~\ref{fig:Fig4}(a) we show the integrated intensity of all four domains (black squares) that measures the order parameter of the HM phase as a function of temperature (see {\it SI Text} on how the intensities were integrated). The magnetic intensity features a sharp drop of about two orders of magnitude at $T_c$, showing that the phase transition is indeed first-order. For comparison we have also plotted the intensity of the critical fluctuations just above the transition (blue circles) demonstrating that the magnetic intensity of the Bragg peaks at low temperature measuring the static HM order parameter is entirely recovered above $T_c$ in the form of critical fluctuations uniformly distributed on a sphere in momentum space.

For a quantitative analysis of the SANS data we have performed fits of the radial $\mathbf{q}$-scans shown in Fig.~\ref{fig:Fig3}. Importantly, we find that the magnetic intensity at $T>T_c$ appearing on the sphere in momentum space can be quantitatively explained in terms of critical HM fluctuations. The radial $\mathbf{q}$-scans are well accounted for [see Fig.~\ref{fig:Fig3}(a)-(d)] by the modified Lorentzian profile as derived by Grigoriev \textit{et al.} \cite{Grigoriev:05}
\begin{align}
 \frac{{\rm d} \sigma({\bf q})}{{\rm d} \Omega}&=
 \mathcal{A}
 \frac{k_B T}{\left((q+Q)^2 + \kappa^2\right)}
 \times\nonumber\\
 &\frac{Q^2 + q^2 + \kappa^2}{(q-Q)^2 + \kappa^2 +
 \alpha_{\rm cub}^2 Q^2
 (\hat{q}_x^4 + \hat{q}_y^4 + \hat{q}_z^4 -1/3)}.
  \label{eq:mnsi_ring_scattering}
\end{align}

\begin{table}[th!]
\caption{Summary of the fit parameters obtained by fitting the inverse correlation length $\kappa(T)$ determined via SANS [see Fig.~\ref{fig:Fig4}(c)] and the magnetic susceptibility $\chi(T)$ [see Fig.~\ref{fig:Fig5}(a)] by means of Eq.~\eqref{eq:kappa_bra} and Eq.~\eqref{eq:chi}, respectively.}\label{tab:Tab1}
\begin{tabular}{@{\vrule height 10.5pt depth4pt  width0pt}p{3.5cm}llll}
Parameter & Symbol & (Unit) & fit of $\kappa(T)$ & fit of $\chi(T)$ \\\hline
inverse Ginzburg length & $\kappa_{\rm Gi}$ & (\AA$^{-1}$) & 0.019(4) & 0.018(3) \\
mean field temperature & $T_{\rm MF}$ & (K) &  30.6 (1.0) & 30.5(5) \\
& $T_0$ & (K) & 1.1(3) & 1.3(2) \\
magnitude of magnetic susceptibility & $\chi_0$ & (1) & $-$ & 0.27(4) \\\hline
\end{tabular}
\end{table}
Here $\mathbf{q}$ is the reduced wave vector with $q = |\bf q|$ measured from the neighboring reciprocal lattice vector, $\alpha_{\rm cub}$ measures the cubic anisotropy, $k_B$ is the Boltzmann constant, and $\mathcal{A}$ is a proportionality factor that depends e.g. on the magnetic form factor of the Mn ions. Eq.~\eqref{eq:mnsi_ring_scattering} describes the intensity of critical magnetic fluctuations with inverse correlation length $\kappa= 2\pi/\xi$ emerging on a sphere with radius $Q$ in reciprocal space. However, depending on the magnitude of the magnetic anisotropy $\alpha_{\rm cub}$ the sphere will have shallow maxima along the $\langle 111\rangle$ directions due to the cubic invariant $(\hat{q}_x^4 + \hat{q}_y^4 + \hat{q}_z^4 -1/3)$, where $\hat q = {\bf q}/q$.

We employed simultaneous fits of $\mathbf{q}$-scans along the $[111]$, $[11\bar{1}]$, $[001]$ and $[110]$ directions to determine the temperature dependence of $\alpha_{\rm cub}$. This is in contrast to Ref.~\cite{Grigoriev:05} where $\alpha_{\rm cub}$ was fixed, resulting in an inverse correlation length $\kappa$ depending on the direction close to $T_c$ in contradiction to Eq.~\eqref{eq:mnsi_ring_scattering} [cf. red crosses and triangles in Fig.~\ref{fig:Fig4}(d)]. Our fits describe the intensity observed on the entire sphere (not only along the fitted directions) remarkably well for all observed temperatures as shown in the {\it SI Text}. For $T=T_c+0.1$~K and $T_c+0.2$~K the fits improve significantly by adding a Gaussian profile to Eq.~\eqref{eq:mnsi_ring_scattering} for the scans parallel to $[111]$ and $[11\bar{1}]$ indicating the presence of a tiny fraction of HM order above $T_c$. This suggests that remanent droplets of the HM phase survive above $T_c$ as expected for a first-order transition \cite{Binder:87}.

The results of the fits are summarized in Figs.~\ref{fig:Fig4}(b)-(d). $Q$ does not show any pronounced anomaly close to $T_c$ and keeps increasing as function of temperature with about the same rate as observed in the HM phase [Fig.~\ref{fig:Fig4}(b)]. In the HM phase $Q$ was determined via fits of the helical satellites using a Gaussian profile (see {\it SI Text}). At $T_c$, we find $Q = 0.039$~\AA$^{-1}$, yielding a pitch length $\xi_{\rm DM} = 2\pi/Q = 160$~\AA.

The cubic anisotropy $\alpha_{\rm cub}$ is shown in Fig.~\ref{fig:Fig4}(c); whereas it is negligible at higher temperatures, it reaches a maximum value of $\alpha_{\rm cub}^2 \approx 0.023$ at $T_c$.
This gives a cubic length scale $\xi_{\rm cub} = 2\pi/(\alpha_{\rm cub} Q) \approx 1040$~\AA. As explained above, the two lengths $\xi_{\rm DM}$ and $\xi_{\rm cub}$ give rise to crossovers in the character of the critical fluctuations, see Fig.~\ref{fig:Fig1}(a). Whether the latter cubic crossover develops or not depends on the size of the correlation length at the first-order transition.

The inverse correlation length $\kappa$ is shown in Fig.~\ref{fig:Fig4}(d) and is well in agreement with previously reported results \cite{Grigoriev:05}. It assumes a finite value at the transition that is given by $\xi_c = 2\pi/\kappa_c \approx 1260$~\AA\, so that it is in fact of similar magnitude as the cubic length, $\xi_{\rm cub}$. This implies that the first-order transition takes place before the cubic anisotropies have fully developed clearly disfavoring a Bak-Jensen scenario.

This is borne out by the value of the Ginzburg length which is at the origin of the peculiar temperature dependence of $\kappa$ as we demonstrate in the following. The temperature dependence of $\kappa$ was previously fitted with an ad-hoc two-stage power-law dependence \cite{Grigoriev:11}. We find, however, that it can be naturally explained in terms of a Brazovskii renormalization. If the singular fluctuation corrections are taken into account self-consistently, one obtains the Brazovskii equation for the inverse correlation length \cite{Brazovski:75} (see {\it SI Text} for details)
\begin{align} \label{BrazKappa0}
\kappa^2 = \kappa_{\rm MF}^2 + \frac{\kappa_{\rm Gi}^3}{\kappa}
\end{align}
where $\kappa_{\rm MF} \propto T-T_{\rm MF}$ measures the distance to the mean-field transition temperature $T_{\rm MF}$, and $\kappa_{\rm Gi}$ can be identified with the inverse Ginzburg length, $\xi_{\rm Gi} = 2\pi/\kappa_{\rm Gi}$. The explicit solution to this cubic equation reads
\begin{equation}
\kappa(T) = \kappa_{\rm Gi} \sqrt{\frac{(\tau + ( 1 - \tau^3 + \sqrt{ 1 - 2\tau^3} )^{1/3})^2}{2^{1/3}(1 - \tau^3 + (\sqrt{1 - 2\tau^3})^{1/3})}},
\label{eq:kappa_bra}
\end{equation}
where $\tau = (2^{1/3}/3) \kappa_{\rm MF}^2/\kappa_{\rm Gi}^2 \equiv (T-T_{\rm MF})/T_0$. As shown in Fig.~\ref{fig:Fig4}(d) (blue solid line) the inverse correlation length that was experimentally determined from our SANS data in the temperature range between $T_c$ and $30.5$~K is perfectly described by the Brazovskii equation~\eqref{eq:kappa_bra} with fit values listed in Table~\ref{tab:Tab1}.

The resulting Ginzburg length amounts to $\xi_{\rm Gi} \approx 330$~\AA\, demonstrating that the three length scales obey $\xi_{\rm DM} < \xi_{\rm Gi} < \xi_{\rm cub}$, see Table~\ref{tab:Tab2} and also Fig.~\ref{fig:Fig4}(e). In turn, the Brazovskii scenario does indeed hold.
At large temperatures $T\gg T_{\rm MF}$, i.e., $\xi \ll \xi_{\rm Gi}$, the system is disordered already on the mean-field level, and Eq.~\eqref{eq:kappa_bra} recovers the mean-field behavior $\kappa \approx \kappa_{\rm MF}$, as shown by the black solid line in Fig.~\ref{fig:Fig4}(c). However, for $T \approx T_{\rm MF}$, i.e., $\xi \approx \xi_{\rm Gi}$, the strong one-dimensional singularity of the interaction correction prevents the correlation length to become infinite and thus impedes the condensation of long-range order. Consequently, the transition temperature is suppressed by $\Delta T = T_{\rm MF} - T_c \approx 1.5$~K, see Table \ref{tab:Tab1} and Fig.~\ref{fig:Fig4}(e), giving rise to a fluctuation-disordered regime for $T_c < T < T_{\rm MF}$ [blue shaded region in Figs.~\ref{fig:Fig4}(d) and (e)]. Finally, the fluctuations trigger a first-order transition into the HM state at a critical $\tau_c < 0$ with a numerical value of order one, $|\tau_c| \sim \mathcal{O}(1)$, whose description is beyond Eq.~\eqref{BrazKappa0} but is explained in detail in the {\it SI Text}.

\begin{figure*}[th!]
\includegraphics[width=1.0\textwidth]{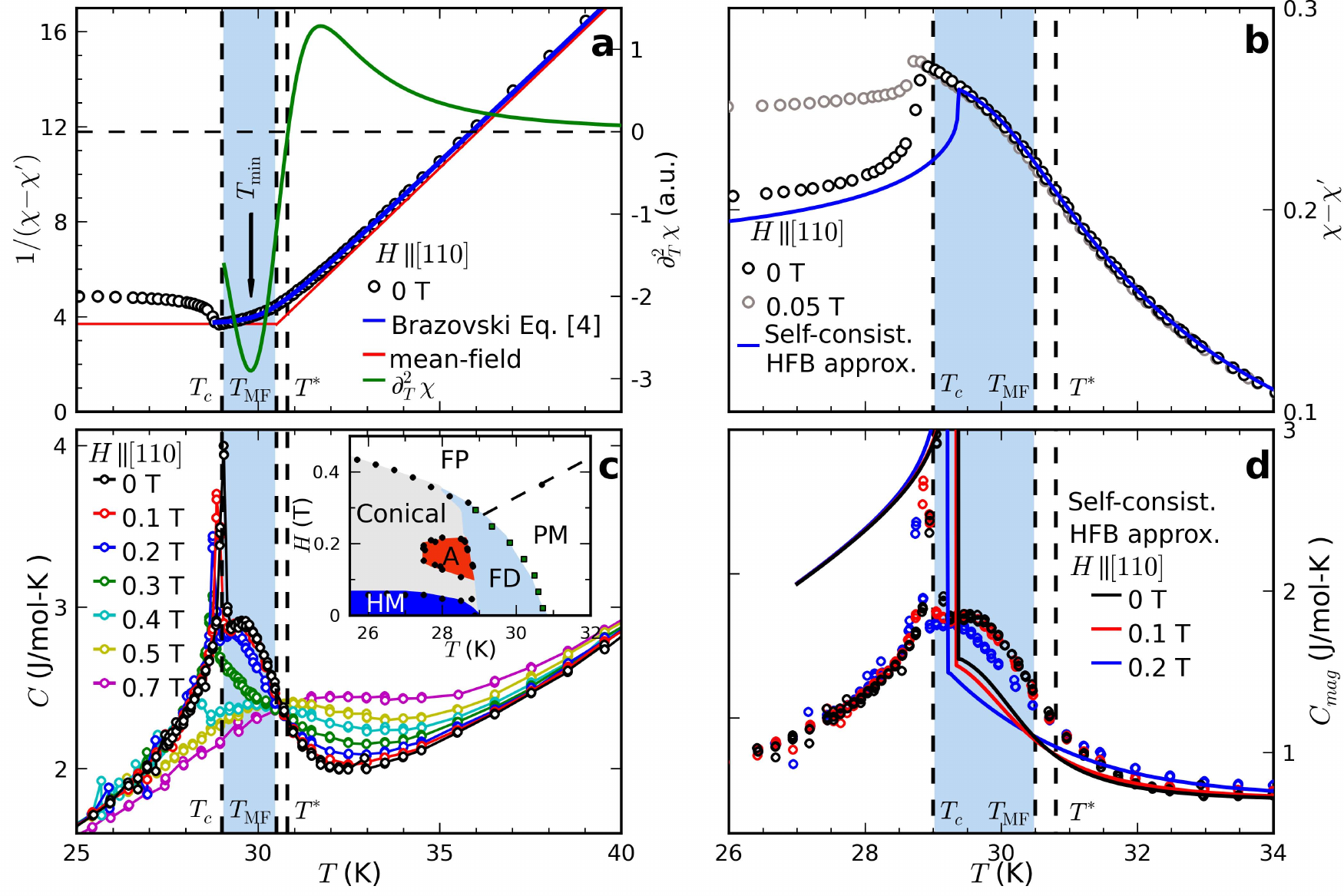}
\caption{ Signatures of the Brazovskii scenario in MnSi as reflected in the thermodynamic observables. (a) The inverse ac-susceptibility $(\chi-\chi')^{-1}$ in zero magnetic field vs. temperature $T$. Here $\chi'$ is a small contribution to $\chi$ that depends on exact position of the sample in the measurement setup. The blue solid line is a fit of the magnetic susceptibility to Eq.~\eqref{eq:chi}. $T_c=29$~K has been determined as the minimum in $\chi^{-1}$. The red solid line shows the (field-cooled) magnetic susceptibility in the mean-field approximation. The green solid line displays $\partial^2_T \chi(T)$ indicating the position of a turning point in $\chi(T)$ ($\partial^2_T \chi(T)=0$, horizontal dashed line) at $T^\ast\approx30.8$~K. $T_{\rm min}\approx29.8$~K denotes the minimum in $\partial^2_T \chi(T)$. (b) The magnetic susceptibility $\chi-\chi'$ vs. $T$ measured for magnetic fields $H=0$ and $0.05$~T, respectively. The blue solid line denotes the magnetic susceptibility calculated via self-consistent Hartree-Fock-Brazovskii (HFB) theory (see text). (c) The specific heat $C$ vs. $T$ for different magnetic fields $H$. Here $T_{c}$ and $T^\ast$ indicate the position of the sharp first-order like feature and a Vollhardt invariance for which $\partial_H C(T)=0$, respectively. We note that the Vollhardt invariance coincides with the turning point in $\chi$. The solid lines are guides to the eye. The inset shows the magnetic phase diagram of MnSi as determined via $\chi$ with $H$ applied along $[110]$. The green squares are the positions of the turning point in $\chi(T)$ that coincide with the crossing point in $C$ in the limit $H\to 0$. HM, FP, FD, and PM denote the helimagnetic, the field-polarized, the fluctuation-disordered paramagnetic and the mean-field disordered paramagnetic regimes, respectively. A labels the A-Phase that shows the skyrmion lattice \cite{Muehlbauer:09}. (d) The magnetic contribution to the specific heat $C_{\rm mag}$ for three magnetic fields $H$ vs. $T$, obtained by subtracting the electronic and lattice parts (see text). The solid lines represent calculations for $C_{\rm mag}$ according to self-consistent HFB theory. Thermodynamics does not show clear signatures at the temperature $T_{\rm MF}$ itself where $\xi \approx \xi_{\rm Gi}$ (cf. Fig.~\ref{fig:Fig4}), but rather features in $\chi$ and $C$ at $T^* \approx 30.8$~K whose value is determined by the ratio $\xi_{\rm Gi}/\xi_{\rm DM}$ (see text).}\label{fig:Fig5}
\end{figure*}

\begin{table}[th!]
\caption{Estimate of various length scales close to the helimagnetic transition of MnSi, see text}\label{tab:Tab2}
\begin{tabular}{@{\vrule height 10.5pt depth4pt  width0pt}lrl}
\hline
chiral DM length & $\xi_{\rm DM}$&$\approx 160$ \AA\\
Ginzburg length & $\xi_{\rm Gi}$&$\approx 330$ \AA\\
cubic anisotropy length & $\xi_{\rm cub}$&$\approx 1040$ \AA\\
correlation length at $T_c$& $\xi_c$&$\approx 1260$ \AA\\
\hline
\end{tabular}
\end{table}

\section{Critical thermodynamics of MnSi}

We now turn to an analysis of the thermodynamic measurements. The longitudinal magnetic susceptibility for temperatures $T>T_c$ and zero magnetic field $H=0$ can also be expressed in terms of the inverse correlation length
\begin{equation}\label{eq:chi}
\chi|_{T>T_c} = \frac{\chi_0}{1+\kappa(T)^2/Q^2},
\end{equation}
where $\chi_0$ is a constant, and $Q$ is the length of the helical propagation vector as before. The inverse of the measured magnetic susceptibility for $H=0$ is shown in Fig.~\ref{fig:Fig5}(a) together with a fit to the Brazovskii formula of Eq.~\eqref{eq:kappa_bra}, which perfectly accounts for the observed $T$-dependence. The fit parameters are given in Table~\ref{tab:Tab1} and agree well with the values extracted from SANS. To highlight the relationship between $\kappa$ and $\chi$, we have added the latter in Fig.~\ref{fig:Fig4}(d) (black circles).

The Brazovskii renormalization of $\kappa$ leads to a striking modification of the temperature dependence of the susceptibility. On the mean-field level, $\kappa_{\rm MF} \propto T-T_{\rm MF}$,  the susceptibility shows Curie-Weiss behavior and would keep increasing as the temperature is lowered [cf. red solid line in Fig.~\ref{fig:Fig5}(a)]. However,
the renormalization of $\kappa$ weakens this increase and, in particular, induces a turning point $\partial_T^2 \chi|_{T=T^*} = 0$ at a temperature $T^*$ = 30.8(1) K, that is slightly larger than $T_{\rm MF}$ (green solid line). To make contact with previous work \cite{Grigoriev:11}, we also note that $\partial_T^2 \chi$ exhibits a local minimum at a temperature $T_{\rm min}=29.8(2)$~K with $T^* > T_{\rm min}$. Within Brazovskii theory, both these temperature scales are determined by the Ginzburg length $\xi_{\rm Gi}$ (or, more precisely, by the ratio $\xi_{\rm Gi}/\xi_{\rm DM}$) and thus originate from the interaction between chiral fluctuations.

As explained in the {\it SI Text}, the self-consistent Hartree-Fock-Brazovskii (HFB) approximation goes beyond Eq.~\eqref{eq:kappa_bra} as it also describes the fluctuation-induced first-order transition and the resulting behavior within the ordered phase. Although a self-consistent Hartree-Fock theory is inadequate for the description of second-order transitions \cite{Baym:77}, it is controlled for Brazovskii systems in the limit of weak interactions, i.e., $\xi_{\rm DM}/\xi_{\rm Gi} \ll 1$. However, as we find $\xi_{\rm DM}/\xi_{\rm Gi} \approx 0.5$ for MnSi, this condition is not well obeyed so that quantitative corrections to HFB theory can in principle be sizable.

In Fig.~\ref{fig:Fig5}(b) a comparison between experiment and theory is shown for the zero-field cooled (ZFC) magnetic susceptibility. The magnetic susceptibility within the ordered phase, $T<T_c$, is anisotropic and thus depends on the domain population. In case of zero-field cooling where all $\langle 111 \rangle$ domains are equally populated this anisotropy results in the decrease of $\chi$ for $T < T_c$  giving rise to the peculiar hooked shape that is nicely reproduced by the HFB approximation. The value of $T_c$, however, is slightly overestimated by theory and the saturation value of the ZFC susceptibility within the ordered phase differs by approximately $10\%$. A small field tends to align ${\bf Q}$ with the field direction resulting in an enhancement of the susceptibility at $H=0.05$T in Fig. 5(b) as compared to zero field.


The specific heat $C$ of MnSi as measured for different magnetic fields $H$ is shown in Fig.~\ref{fig:Fig5}(c). At zero field a sharp first-order spike is observed at $T_c$. With increasing $H$ it shifts to slightly lower temperatures, while simultaneously its magnitude is suppressed, well in agreement with the known phase diagram shown in the inset of Fig.~\ref{fig:Fig5}(c). The first-order peak is accompanied by a broad shoulder with a slope that decreases with $H$ resulting in a characteristic crossing point at $T^*$~$=$~30.8~K. At large fields $H> H_{c2}\approx 0.55$~T \cite{Bauer:10}, the HM order is suppressed and the first-order transition disappears while the shoulder develops into a broad feature indicating the crossover from a PM into a field-polarized regime.

Crossing points of the specific heat in general have been discussed by Vollhardt \cite{Vollhardt:97} who pointed out that they are linked to inflection points of a certain conjugate variable by a Maxwell relation. In our case, the crossing point in $C$ is related to an inflection point in the magnetization $M$,
\begin{align}
0 = \partial_H C\big|_{T^*} = T \partial^2_T M \big|_{T^*}  \overset{H \to 0 }{\approx} T H \partial^2_T \chi \big|_{T^*}.
\end{align}
In the limit of small fields where $M = \chi H$, this is equivalent to a turning point of the susceptibility $\chi$ which, as discussed above, is induced by the Brazovskii renormalization of the correlation length. The turning point in $\chi$ and the crossing point in $C$ are thus different manifestations of the same phenomenon. The strong renormalization effects arising from the interaction among chiral fluctuations is therefore also at the origin of the Vollhardt invariance observed in $C$.

Fig.~\ref{fig:Fig5}(d) shows the magnetic contribution to the specific heat $C_{\rm mag}$ that has been obtained by subtracting the electronic and lattice contribution using the values established in Ref. \cite{Bauer:10}. After having fixed all parameters by a fit to the susceptibility, the self-consistent HFB approximation predicts the behavior for $C_{\rm mag}$ as shown by the solid lines. It nicely accounts for the shoulder and the Vollhardt invariance close to $T^*$ and explains the quasi-saturation just above the first-order transition even though the parameter $\xi_{\rm DM}/\xi_{\rm Gi}$, that controls this approximation, is not particularly small. The HFB approximation becomes less accurate with increasing field and at $H=0.2$~T ceases to be reliable (see {\it SI text}). Within the ordered phase, $T<T_c$, the HFB approximation shows a pronounced additive contribution to the specific heat reminiscent of the jump in $C$ on the level of mean-field theory. Interestingly, this offset is however not observed experimentally which might be attributed either to an additional higher order $M^6$-term in the Ginzburg-Landau expansion or to the influence of HM Goldstone modes \cite{Janoschek:10} (see {\it SI text}).

\section{Discussion and Outlook}

In summary, we showed that the transition from the PM to the HM phase in MnSi is driven weakly first-order by fluctuations. The first-order nature of the transition is confirmed by our SANS measurement, and the intensity of the critical fluctuations for $T >T_c$ is found to be practically uniformly distributed on a sphere in momentum space with only small corrections due to cubic anisotropies. This identifies the Brazovskii scenario as the driving mechanism for the fluctuation-induced first-order transition and rules out the scenario proposed by Bak and Jensen.
This is corroborated by the observed suppression of the transition temperature and, in particular, by the temperature dependence of the measured correlation length that can be quantitatively described by the lowest-order self-consistent Brazovskii approximation. We discussed the relation between the critical fluctuations and thermodynamic quantities, and, especially, explained how the Brazovskii renormalization of the correlation length induces an inflection point in the magnetic susceptibility and a Vollhardt crossing point in the specific heat. The provided evidence for these thermodynamic signatures convincingly rules out alternative explanations in terms of intermediate skyrmion liquid phases as proposed in Refs.~\cite{roessler:06,Pappas:09,Pappas:11, Hamann:11}.

Our study also offers interesting perspectives for future research. An obvious question to be addressed is the evolution of the Brazovskii fluctuations with increasing pressure. The transition temperature of MnSi can be suppressed by applying a pressure of $p \approx 15$~ kbar accompanied by the emergence of partial magnetic order and an unusual extended non-Fermi liquid phase \cite{Pfleiderer:01,Pfleiderer:04}. The possible relation of these phenomena to a quantum version of Brazovskii theory \cite{Schmalian:04} is an interesting open question. Furthermore, as the magnetic field increases the first-order nature of the transition weakens and it eventually becomes second order resulting in a tricritical point that is to be investigated [cf. inset of Fig.~\ref{fig:Fig5}(c)].

As our arguments are derived from basic symmetry principles we expect that cubic DM-HMs with weak anisotropies like the B20 compounds generically exhibit a fluctuation-induced first-order transition of Brazvoskii type. However, one may raise the question whether such transitions also prevail in DM-HMs with lower symmetry, such as Ba$_2$CuGe$_2$O$_7$ (tetragonal) \cite{Zheludev:99} -- for which critical magnetic fluctuations similar to the ones in MnSi have been already observed \cite{Basti}, Ba$_3$NbFe$_3$Si$_2$O$_{14}$ (trigonal) \cite{Marty:09}, and NdFe$_3$(BO$_3$)$_4$ (trigonal) \cite{Janoschek:10a}. Notably, DM-HMs with lower symmetry are relevant for multiferroics \cite{Cheong:07} and magnetic sur- and interfaces \cite{Bogdanov:01, Bode:07, Heide:08} that are promising systems for memory/storage devices.

Our work identifies the cubic DM-HMs as a class of systems that realize fluctuation-induced first-order transitions of the Brazvoskii type, whose unequivocal experimental confirmation has proven
elusive in the past. This paves the way for a more detailed experimental study of such transitions and their properties, e.g., the nucleation of critical HM droplets \cite{Hohenberg:95}.

\begin{acknowledgments}
The neutron scattering experiments for this work have been performed at the very cold neutron multipurpose beamline MIRA\cite{Georgii:07}, situated at the Forschungsneutronenquelle Heinz Maier-Leibnitz (FRM II). We acknowledge financial support through DFG-FOR960 (Quantum phase transitions), DFG-SFB608 (Complex transition metal compounds with spin and charge degrees of freedom and disorder), DFG-TRR80 (From electronic correlations to functionality) , ERC AdG (291079). MJ and AB would like to thank the Alexander von Humboldt foundation and TUM graduate school for financial support, respectively. Further we are grateful to S. M{\"u}hlbauer and R. Schwikowski for experimental support and C. Pappas, W. H{\"a}u{\ss}ler and, in particular, A. Rosch for useful discussions.
\end{acknowledgments}

\end{bibunit}

\begin{bibunit}
\balancecolsandclearpage
\section*{SUPPORTING INFORMATION}

\setcounter{section}{0}

In the first section of this supporting information details of the analysis of the small-angle neutron scattering (SANS) data are presented. The remaining sections explain the theoretical analysis. In the second section the Ginzburg-Landau theory for cubic helimagnets (HM) is introduced and the mean-field properties are reviewed. The third section discusses the application of the Hartree-Fock-Brazovskii (HFB) approximation to helimagnets, and the fourth section presents estimates of the model parameters and energy scales for MnSi.

\section{Analysis of the SANS data}
\subsection{Integration of magnetic intensities}

\begin{figure}[bh!]
\includegraphics[width=0.5\textwidth]{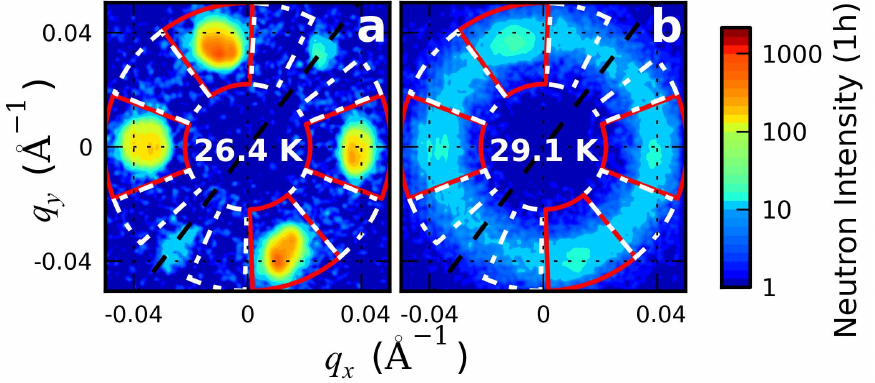}
\caption{The angular sections used for the integration of the magnetic intensities observed on the magnetic satellite peaks associated with the helimagnetic order in MnSi (red solid lines) and magnetic fluctuations observed on a ring (sphere) above $T_c$~$=$~29~K (white broken lines) are illustrated. Here (a) and (b) show examples of the intensity in the helimagnetic phase below $T_c$ and the fluctuations above $T_c$, respectively. The black dashed line indicates a direction along which magnetic Bragg peaks are observed due to double scattering, and which has therefore not been included in the integration (see text for details).}\label{fig:FigS1}
\end{figure}

We have separated the intensities belonging to the helical Bragg satellites and the magnetic fluctuations observed on the sphere by carefully choosing appropriate integration regions as shown in Fig.~\ref{fig:FigS1}. Here the segments with the red solid and white broken border were used to integrate the intensities for the helical satellites and the ring, respectively.

The segments for the ring have been broken up into six pieces in order to remove the intensity of the peaks that develop due to double scattering in the sample (see black broken line in Fig.~\ref{fig:FigS1}). The full integrated intensity of the observed ring was consequently obtained by assuming that the intensity is isotropically distributed on the ring and scaling the intensity by $\pi/l_R$, where $l_R$ is the combined arc length of all six segments. The intensity on the whole sphere can be estimated from the intensity of the observed ring by considering the angular resolution of the experiment $\delta\omega$~=~4.5$^\circ$~=~0.025$\pi$ (FWHM), that was extracted from Gaussian fits to rocking scans over the magnetic satellites (cf. Fig.~\ref{fig:Fig2} in {\it main text}). We note that cold triple axis measurements on this sample show that the magnetic satellite reflections are narrow ($<$ 0.1$^\circ$) \cite{Janoschek:10S}, thus demonstrating that the measured peak width in our experiment is resolution limited, and the angular resolution in a rocking scan is characterized by $\delta\omega$. The surface segment of the sphere covered by the ring at a single position of $\omega$ is $S = 2\int_0^\pi\sin(\theta) d\theta \int_0^{\delta\omega}d\varphi = 4\delta\omega$, indicating that the integrated intensity of the sphere should approximately amount to $I_{Sphere}=\frac{4\pi}{4\delta\omega}I_R=40I_R$, where $I_R$ is the intensity of the ring. $I_{Sphere}$ is plotted as the blue circles in Fig.~\ref{fig:Fig4}(a) in the {\it main text}.

The integrated intensity on the helical satellites $I_{Sat}^{1,2}$ belonging to domains $\mathbf{Q}_1$ and $\mathbf{Q}_2$ has been calculated by subtracting the contribution from the ring to the intensities in the red solid annular segments ($I_{RS}$): $I_{Sat} = I_{RS} - I_R l_{RS}/\pi$, where $l_{RS}$ is the combined arc length of all four red segments. The integrated intensity for all four helical domains $I_{Sat}^{all}$ has been consequently obtained by multiplying $I_{Sat}^{1,2}$ by a factor 2 based on the knowledge that all four domains are generally equally populated \cite{Janoschek:10S}. $I_{Sat}^{all}$ is plotted as the black squares in Fig.~\ref{fig:Fig4}(a) in the {\it main text}.

\subsection{Fits of the helimagnetic Bragg peaks below $T_c$}

\begin{figure}[th!]
\includegraphics[width=.5\textwidth]{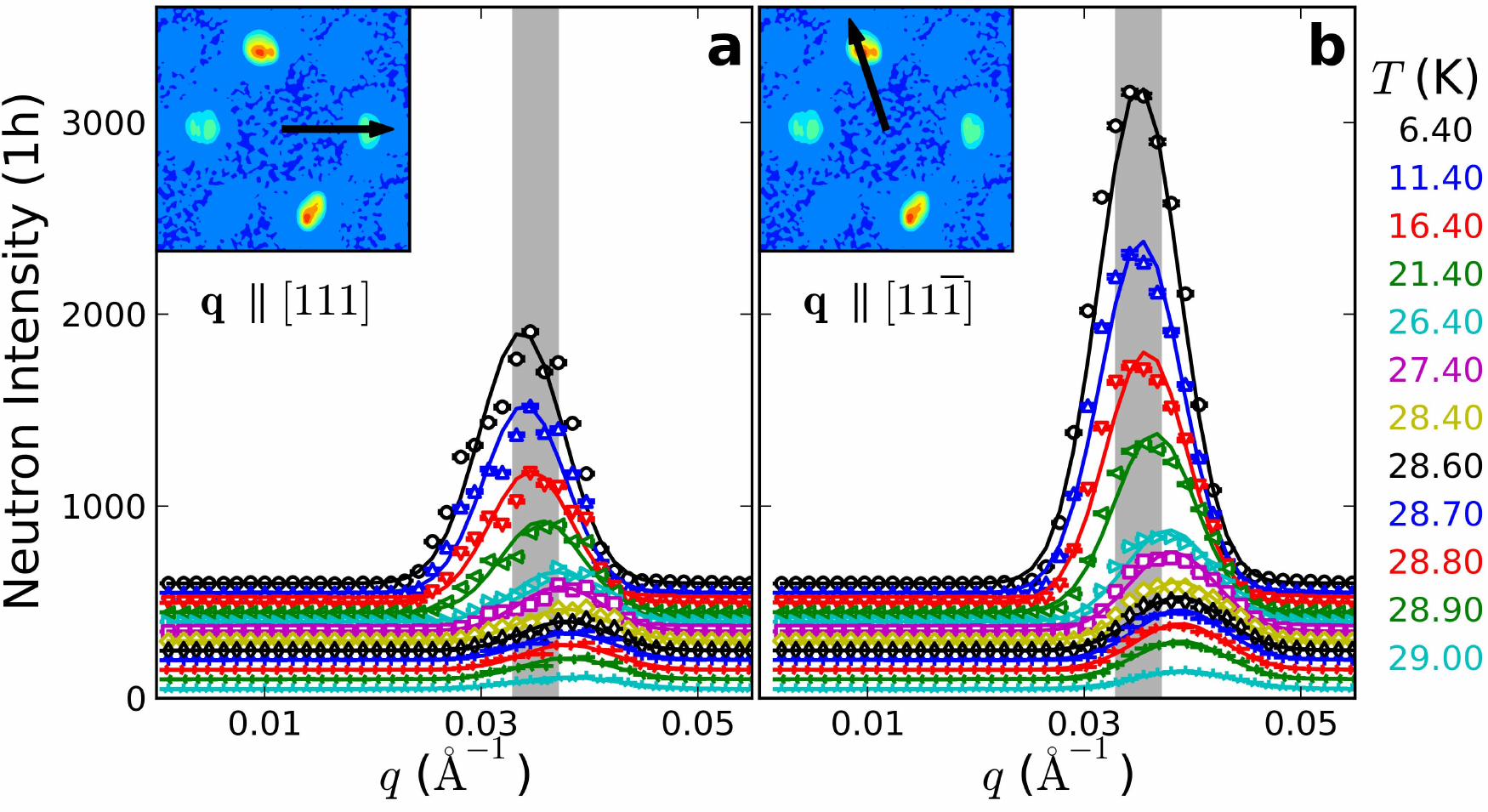}
\caption{Radial $\mathbf{q}$-scans through the helimagnetic Bragg peaks below $T_c$~$=$~29 K are shown for all measured temperatures. The inset in each panel shows the corresponding direction of the scan in reciprocal space. The $\mathbf{q}$-scans were extracted from the two-dimensional detector images in Fig.~\ref{fig:Fig1} in the {\it main text} by performing radial bins with an azimuthal width corresponding to the experimental resolution. The solid lines are fits to Gaussian lineshapes (see text). The grey shaded region shown in each panel denotes the experimental resolution.}\label{fig:FigS3}
\end{figure}

To obtain the temperature-dependence of the magnitude of the helimagnetic propagation vector $Q$ we have performed fits of the radial $\mathbf{q}$-scans through the helimagnetic Bragg peaks shown in Fig.~\ref{fig:FigS3}. The peaks were fitted with a Gaussian lineshape where the peak position determines the distance to the center of the Brillouin zone and therefore the magnitude of $Q$. The temperature dependence of $Q$ is shown in Fig.~\ref{fig:Fig4}(b) in the {\it main text}.

\subsection{Calculation of the intensity of magnetic fluctuations}

\begin{figure*}[th!]
\includegraphics[width=1.0\textwidth]{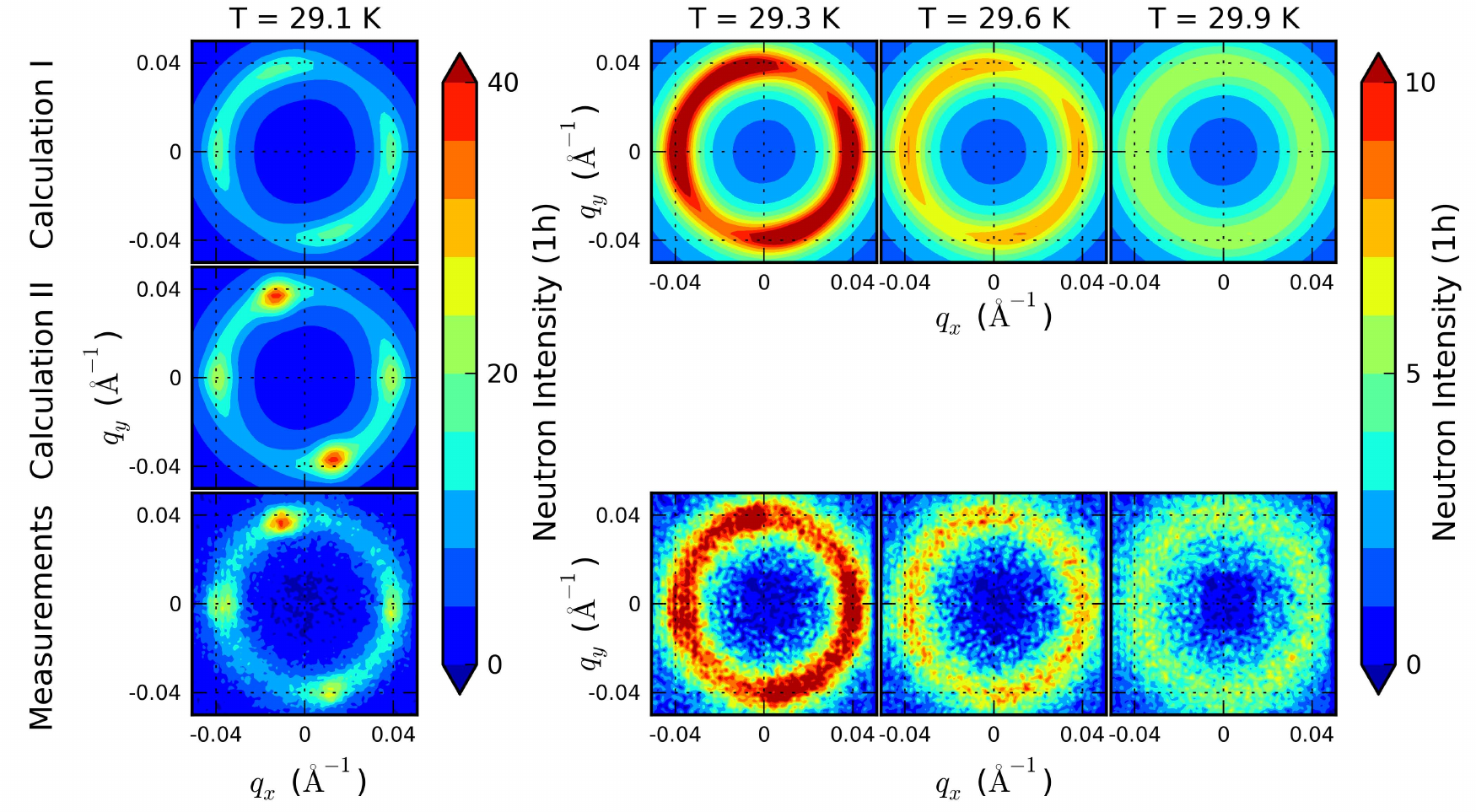}
\caption{A comparison between the calculated (two upper rows) and measured intensities (bottom row) of the magnetic fluctuations above $T_c$~$=$~29~K. Here the calculated intensities where obtained by using Eq.~\eqref{eq:mnsi_ring_scattering} in the main manuscript that describes magnetic fluctuations arising at the phase transition between the helimagnetic and the paramagnetic state in a Dzyaloshinskii-Moriya helimagnet. For the calculation the fit parameters (s. Fig.~3 in {\it main text}) obtained by fits of radial $\mathbf{q}$-scans in the four main directions [111], [11$\bar{1}$], [001]  and [110] (s. Fig.~4(b) in {\it main text}) have been employed. For calculation II additional Gaussian profiles have been added at the position of the helical Bragg satellites in the helimagnetic phase for $T$~$=$~29.1~K (see text for details).}\label{fig:FigS2}
\end{figure*}

The fits of the magnetic fluctuations above $T_c$ to Eq.~\eqref{eq:mnsi_ring_scattering} in the {\it main text} have been carried out by doing combined fits of radial $\mathbf{q}$-scans in the four main directions [111], [11$\bar{1}$], [001] and [110] that characterize the magnetic cubic anisotropy on the ring (sphere). However, Eq.~\eqref{eq:mnsi_ring_scattering} can be used to obtain the entire intensity of the fluctuations in all directions on the sphere. In order to verify how well
our fit describes the intensity on the entire sphere we have performed a calculation of these intensities based on Eq.~\eqref{eq:mnsi_ring_scattering} using the corresponding parameters obtained in the fit (see Fig.~\ref{fig:Fig4} in {\it main text} for the fit parameters). As illustrated in Fig.~\ref{fig:FigS2} for four representative temperatures the calculated intensity of the magnetic fluctuations is in excellent agreement with the experiments. The only temperature where major disagreement is observed is $T$~$=$~29.1~K just above $T_c$ (cf. calculation I in Fig.~\ref{fig:FigS2}). As reported in the main manuscript between $T_c$ and 29.2~K an additional Gaussian shaped intensity component was observed at the position of the helical satellites below $T_c$ indicating that remanent droplets of helimagnetic order survive above $T_c$. Adding these Gaussian components according to the fit parameters obtained via the fit of radial $\mathbf{q}$ (assuming isotropic Gaussian linewidth) for the calculation for $T$~$=$~29.1~K results in excellent agreement for this temperature as well (cf. calculation II in Fig.~\ref{fig:FigS2}).

\section{Ginzburg-Landau theory of chiral magnets}
\label{GLtheory}

The classical Ginzburg-Landau theory of a chiral magnet is given by the free energy functional $\mathcal{F} = \int d^3 x\, f$ with the density $f = f_0 + f_{\rm cub}$ where \cite{Bak:80S,Nakanishi:80}
\begin{align} \label{model}
f_0 = \frac{1}{2} \vect\phi (r - J \nabla^2) \vect \phi + D \vect \phi (\nabla \times \vect \phi) + \frac{u}{4!} (\vect \phi^2)^2
- \gamma \vect \phi \vect H.
\end{align}
We choose dimensionless units for the three component order parameter field $\vect \phi$, which we normalize such that the coupling to the magnetic field $\vect H$ is just given by the magnetization density $\gamma = \mu_B/$f.u. corresponding to a single Bohr magneton per formula unit that is f.u. $= 24.018$ \AA$^3$ for MnSi. The absence of inversion symmetry allows for the Dzyaloshinskii-Moriya (DM) interaction $D$ that couples internal magnetic space and real space and is proportional to the strength of the spin-orbit coupling $\lambda_{\rm SO}$, $D \sim \mathcal{O}(\lambda_{\rm SO})$, which is parametrically small in MnSi.
The terms of second and higher order in spin-orbit coupling are contained in $f_{\rm cub}$ that, in particular, break the rotation symmetry of $f_0$ present at $\vect H=0$ due to cubic anisotropies. We restrict ourselves in the following to a single, representative term
\begin{align} \label{modelCubAn}
f_{\rm cub} = \frac{J_{\rm cub}}{2} \left((\partial_x \phi_x)^2 + (\partial_y \phi_y)^2+(\partial_z \phi_z)^2 \right) + ...
\end{align}
where the coupling constant $J_{\rm cub} \sim \mathcal{O}(\lambda_{\rm SO}^2)$.

\subsection{Mean-field theory}

The benchmark in the following will be the mean-field approximation that we review here. The Ansatz for a single conical helix reads
\begin{align} \label{MFAnsatz}
\vect \phi_{\rm hel}(\vect r) = \hat \phi_0 \phi_0 + \Psi_{\rm hel} \hat e^- e^{i \vect Q \vect r} + \Psi^*_{\rm hel} \hat e^+ e^{-i \vect Q \vect r},
\end{align}
where $\phi_0$ is the homogeneous magnetization, $\Psi_{\rm hel}$ is the complex amplitude of the helical order characterized by the pitch vector $\vect Q$. The vectors are given by $\hat e^\pm = (\hat e_1 \pm i \hat e_2)/\sqrt{2}$ with the normalized dreibein $\hat e_1 \times \hat e_2 = \hat Q$ where $\hat Q = \vect Q/Q$. Evaluating the energy density $f$ with this Ansatz one obtains the mean-field potential $\mathcal{V} = \mathcal{V}_0 + \mathcal{V}_{\rm cub}$ where
\begin{align} \label{PotMF}
\mathcal{V}_0 =&
\frac{r}{2} \phi_0^2 + (r + J Q^2 - 2 D Q) |\Psi_{\rm hel}|^2
\\\nn&+ \frac{u}{4!} (\phi_0^2 + 2 |\Psi_{\rm hel}|^2)^2 - \gamma  \phi_0 H,
\end{align}
determines the strength of the amplitudes. The second contribution $\mathcal{V}_{\rm cub}$ reflects the competition between the magnetic field and the cubic anisotropies to orient the pitch vector $\vect Q$ and the homogeneous magnetization $\hat \phi_0$,%
\begin{align} \label{Potcub}
\mathcal{V}_{\rm cub} &= \gamma  \phi_0 H ( 1 - \hat \phi_0 \hat H)+ \frac{u}{4!} |\Psi_{\rm hel}|^2 \phi_0^2 (\hat Q \times \hat \phi_0)^2
\\\nn
&+ \frac{J_{\rm cub}}{2} |\Psi_{\rm hel}|^2 Q^2 (1-
(\hat Q_x^4 + \hat Q_y^4 +\hat Q_z^4)
).
\end{align}

In the paramagnetic phase $|\Psi_{\rm hel}| = 0$, one has $\hat \phi_0 = \hat H$ and minimization $\partial_{\phi_0}\mathcal{V} = 0$
results in the magnetic equation of state,
\begin{align} \label{mEoS}
r \phi_0 + \frac{u}{3!} \phi_0^3 = \gamma H.
\end{align}
On the other hand, at zero field $H=0$ the homogeneous magnetization vanishes $\phi_0 = 0$. From Eq.~\eqref{Potcub} follows that in the helimagnetic ordered phase $|\Psi_{\rm hel}| > 0$, even a tiny cubic anisotropy $J_{\rm cub}$ is sufficient to orient the pitch vector along a cubic symmetry direction. We assume $J_{\rm cub} < 0$ as this prefers the $\langle 111 \rangle$ direction as observed in MnSi \cite{Bak:80S,Nakanishi:80}. Minimization with respect to the pitch length then yields $Q = D/J_{\rm eff}$ where $J_{\rm eff}=J + J_{\rm cub}/3$, and the effective potential for $H=0$ then reduces to
\begin{align} \label{EffPot}
\mathcal{V} =
\delta |\Psi_{\rm hel}|^2 + \frac{u}{3!} |\Psi_{\rm hel}|^4,
\end{align}
where we introduced the helimagnetic tuning parameter
\begin{align} \label{TuningParam}
\delta = (r - J_{\rm eff} Q^2) \approx r- J Q^2.
\end{align}
In the last equation we neglected small corrections of relative order $|J_{\rm cub}|/J \ll 1$. Helimagnetic order develops in the form of a second-order mean-field transition if the  tuning parameter $\delta$ becomes negative at $H=0$, i.e., if the parameter $r$ is reduced below the DM energy density $r \leq J Q^2$. Minimization of the effective potential then yields for the amplitude $|\Psi_{\rm hel}|^2 = - 3 \delta/u$. If the magnetic field $H$ is increased within the helimagnetically ordered phase and points away from $\langle 111\rangle$, the competition between the cubic anisotropy and magnetic energy in $\mathcal{V}_{\rm cub}$ results in a reorientation of the pitch at a critical field $H_{c1}$ whose precise value depends on the orientation of the applied field, $\hat H$, with respect to the crystallographic $\langle 111\rangle$ direction.

\subsection{Fluctuation spectrum and crossover scales}

The fluctuations around the mean-field solution are described by the susceptibility tensor $\chi_{0, ij}^{-1}(\vect r,\vect r') = \delta^2 \mathcal{S}/(\delta \vect \phi_i(\vect r)\delta \vect \phi_j(\vect r'))$ with the action $\mathcal{S}[\vect \phi] = \int d^3 x f/(k_B T)$, which reads explicitly
\begin{align} \label{BareSusc}
&\chi^{-1}_{0,ij}(\vect r,\vect r') =  \frac{1}{k_B T}
\Big[ (r - J \nabla_{\vect r}^2) \delta_{ij} - 2 D \varepsilon_{ijn} \nabla_{\vect r_n}
\\\nn
&+ \frac{u}{3!}  \left(\vect \phi({\vect r})\vect \phi({\vect r}) \delta_{ij} + 2 \vect \phi_i({\vect r}) \vect \phi_j({\vect r})  \right)
\Big]\delta(\vect r - \vect r') +\chi^{-1}_{{\rm cub}\, ij}(\vect r,\vect r') .
\end{align}
where $\vect \phi({\vect r})$ is the mean-field order parameter and $\chi^{-1}_{{\rm cub}}$ is the contribution arising from the cubic anisotropies $f_{\rm cub}$ of Eq.~\eqref{modelCubAn}.

In the paramagnetic phase, $|\Psi_{\rm hel}| = 0$, the Fourier transform of the susceptibility, $\chi_0^{-1}(\vect k, \vect k') = \chi_0^{-1}(\vect k)  \delta_{\vect k, -\vect k'}$, is diagonal in momentum space. As shown by Grigoriev {\it et al} \cite{Grigoriev:05S}, inverting this susceptibility, $\chi_0^{-1}(\vect k)$, for magnetic field $H=0$ and taking into account the cubic anisotropy $J_{\rm cub}$ close to the critical singularity yields the form of the scattering cross section  of Eq.~\eqref{eq:mnsi_ring_scattering} in the {\it main text}, that were used to analyze the neutron scattering data with the identification $\alpha_{\rm cub}^2 = |J_{\rm cub}|/(2J)$.

It is instructive to consider the mean-field fluctuation spectrum $\omega_{0,\vect k}$ that follows from the eigenvalue equation
\begin{align} \label{SuscSpectrum}
k_B T \sum_{\vect k'} \chi_{0,ij}^{-1}(\vect k,\vect k') v_j (-\vect k') =  \omega_{0,\vect k} v_i(\vect k)
\end{align}
with eigenvectors $v_j(\vect k)$ as a function of momentum $\vect k$. This spectrum changes qualitatively as the transition is approached from high temperatures, $|\Psi_{\rm hel}|=0$, and it exhibits a series of crossovers as illustrated in Fig.~\ref{fig:Fig1}(a) in the {\it main text}.
Far above the transition, the spectrum is approximately that of a
ferromagnet above its Curie temperature $\omega_{0,\vect k} \approx r + J k^2$. As the temperature is lowered and $r$ reaches the order of the Dzyaloshinskii-Moriya energy density
\begin{align}
\varepsilon_{\rm DM} = J Q^2
\end{align}
fluctuations start to become soft on a sphere in momentum space with radius $Q$ and the magnetic correlations develop an oscillating component. The corresponding spectrum has three branches $\omega^{\rm cr}_{0,\vect k} < \omega^{\rm 1}_{0,\vect k}  < \omega^{\rm 2}_{0,\vect k}$. The low-energy part has the Brazovskii-form
\begin{align} \label{BrazSpectrum}
 \omega^{\rm cr}_{0,\vect k} = \delta + J (|\vect k| - Q)^2
\end{align}
where $\delta$ is the helimagnetic tuning parameter of Eq.~\eqref{TuningParam}. The dispersion of the remaining two fluctuation branches are $ \omega^{\rm 1}_{0,\vect k} = \delta + J Q^2 + J \vect k^2$ and $\omega^{\rm 2}_{0,\vect k} = \delta + J (|\vect k| + Q)^2$.
Close to criticality as $\delta \to 0^+$, these other modes have still a gap on order the DM interaction energy density $\varepsilon_{\rm DM} = J Q^2$. The correction to the critical spectrum \eqref{BrazSpectrum} arising from cubic anisotropies of Eq.~\eqref{modelCubAn} finally becomes important for values of the tuning parameter on the order of the cubic energy density, $\delta \sim \varepsilon_{\rm cub}$, with
\begin{align} \label{EnergyCub}
\varepsilon_{\rm cub} = \alpha_{\rm cub}^2\, \varepsilon_{\rm DM} = |J_{\rm cub}| Q^2/2.
\end{align}
The energy scales that determine the crossover in the fluctuation spectrum are converted into the length scales $\xi_{\rm DM}$ and $\xi_{\rm cub}$ of
Fig.~\ref{fig:Fig1}(a) in the {\it main text} via $\varepsilon_{\rm DM} = J (2\pi/\xi_{\rm DM})^2$ and $\varepsilon_{\rm cub}= J(2\pi/\xi_{\rm cub})^2$.

\subsection{Mean-field magnetic susceptibility}

For zero momenta the susceptibility matrix \eqref{BareSusc} reduces to the thermodynamic dimensionless magnetic susceptibility via the relation $\chi^{\rm MF}_{{\rm mag},ij} = \frac{\gamma^2 \mu_0}{k_B T} \chi_{0,ij}(0,0)$.

In the paramagnetic phase, $|\Psi_{\rm hel}|=0$, at zero field, $H=0$, one obtains $\chi^{\rm MF}_{{\rm mag},ij} = \delta_{ij}(\gamma^2 \mu_0)/r$.
In the helimagnetically ordered phase, $|\Psi_{\rm hel}| > 0$, on the other hand, the mean-field order parameter $\vect \phi({\vect r})$ carries a finite momentum $\vect Q$ and, as a result, the susceptibtility is non-diagonal in momentum space. Using the Ansatz \eqref{MFAnsatz} we obtain
\begin{align} \label{BareSusc2}
&\chi^{-1}_{0,ij}(\vect k, \vect k') = \chi^{-1}_{0,ij}(\vect k) \delta_{0, \vect k+\vect k'}
\\\nn
&+ \frac{u}{3} \psi^2_{\rm hel} \hat e^-_i \hat e^-_j
\delta_{-2 \vect Q, \vect k + \vect k'}
+ \frac{u}{3} {\psi^{*2}_{\rm hel}} \hat e^+_i \hat e^+_j
\delta_{2 \vect Q, \vect k + \vect k'}
\\\nn
&+ \frac{u}{3} \psi_{\rm hel} \phi_0 (\hat e^-_i \hat \phi_{0 j} + \hat e^-_j \hat \phi_{0 i})
\delta_{-\vect Q, \vect k + \vect k'}
\\\nn
&+\frac{u}{3} \psi^*_{\rm hel} \phi_0 (\hat e^+_i \hat \phi_{0 j} + \hat e^+_j \hat \phi_{0 i})
\delta_{\vect Q, \vect k + \vect k'}.
\end{align}
The part that is diagonal in momenta reads
\begin{align}
&\chi^{-1}_{0,ij}(\vect k) =
\frac{1}{k_B T}\Big[
(r + J k^2) \delta_{ij} - 2 D \epsilon_{ijn} i \vect k_n +
\\\nn
&+ \frac{u}{3!} \phi_0^2 (\delta_{ij} + 2 \hat \phi_{0 i} \hat \phi_{0 j})
+\frac{u}{3} |\psi_{\rm hel}|^2 (2\delta_{ij} - \hat Q_{i} \hat Q_{j})
\Big]+\chi^{-1}_{{\rm cub}\, ij}(\vect k).
\end{align}

The magnetic susceptibility is obtained from the inverse of the generalized matrix $\chi^{-1}_{0,ij}(\vect k, \vect k')$ taken at zero momenta. It is however important to note that $\chi_{0,ij}(0,0) \neq (\chi^{-1}_{0,ij}(0,0))^{-1}$ because the order parameter carries momentum. After inverting the susceptibility matrix, we get for the dimensionless magnetic susceptibility of a single helimagnetic domain $\chi^{\rm MF}_{{\rm mag},ij} = \frac{\gamma^2 \mu_0}{k_B T} \chi_{0,ij}(0,0)$ at zero magnetic field $H=0$, i.e., $\phi_0=0$
\begin{align}
\chi^{\rm MF}_{{\rm mag},ij} = \frac{\gamma^2 \mu_0}{J Q^2} \left(\hat Q_i \hat Q_j + \frac{1 - \delta/(JQ^2)}{1- 2 \delta/(J Q^2)} (\delta_{ij} - \hat Q_i \hat Q_j )\right)
\end{align}
where we neglected corrections arising from cubic anisotropies for simplicity. Moreover, we used the equation of state $|\Psi_{\rm hel}|^2 = - 3\delta/u$ with the tuning parameter $\delta$ of Eq.~\eqref{TuningParam} that is negative in the helimagnetically ordered phase, $\delta < 0$. Close to the transition $\delta \to 0^-$ the magnetic susceptibility becomes isotropic and smoothly connects to the value in the paramagnetic phase. Deep in the ordered phase, $\delta \ll - J Q^2$, the susceptibility is anisotropic. In particular, the longitudinal susceptibility, $\chi_{{\rm mag}} = \chi_{{\rm mag},ij} \hat H_i \hat H_j$, depends on the angle between the magnetic field and the pitch vector, $\hat H\hat Q$,
\begin{align}
\chi^{\rm MF}_{{\rm mag}} = \frac{\gamma^2 \mu_0}{J Q^2} \frac{(1 - \delta/(JQ^2)) + (\hat H\hat Q)^2 (- \delta/(JQ^2))}{1- 2\delta/(JQ^2)}.
\end{align}
The macroscopic magnetic susceptibility $\langle \chi_{{\rm mag}}\rangle$ averages over helimagnetic domains that might possess pitches in different $\langle 111 \rangle$ directions. For field cooling (FC) with $\vect H$ in a $\langle 111 \rangle$ direction, we might expect a single helimagnetic domain with $(\hat H\hat Q)^2 = 1$ so that $\langle \chi^{\rm MF}_{{\rm mag}}\rangle^{\langle 111 \rangle}_{\rm FC} = \gamma^2 \mu_0/(JQ^2)$ attains the maximal possible value independent of $\delta$. This is also the value of the longitudinal susceptibility one expects in the conical phase at a finite field $H>H_{\rm c1}$.
For zero field cooling (ZFC), on the other hand, we can assume an equal distribution of domains so that $\langle (\hat H\hat Q)^2 \rangle = 1/3$ independent of the magnetic field orientation
\begin{align} \label{MagSuscZFC}
\langle \chi^{\rm MF}_{{\rm mag}}\rangle|_{\rm ZFC} = \frac{\gamma^2 \mu_0}{J Q^2} \frac{1 - \frac{4}{3}\delta/(JQ^2)}{1- 2\delta/(JQ^2)} .
\end{align}
As the helimagnetic phase is entered this susceptibility drops from the mean-field value at criticality $\gamma^2 \mu_0/(J Q^2)$ and saturates for $\delta \ll - J Q^2$ to a value that is reduced by a factor of $2/3$.

\section{Brazovskii theory for chiral magnets}

Close to the critical temperature, chiral magnetic fluctuations become very important and have to be treated in a self-consistent manner.
Corrections arising from the interaction of modes with the dispersion of Eq.~\eqref{BrazSpectrum} are
singular as these fluctuations have a quasi one-dimensional character. As the momentum dependence is peaked at a finite momentum along  the radial direction, the density of states %
\begin{align} \label{DOS1d}
\nu(\varepsilon) = \int \frac{d \vect k}{(2\pi)^3} \delta(\varepsilon - \omega^{\rm cr}_{0,\vect k}) \approx \frac{Q^3}{2\pi^2 \sqrt{J Q^2}} \frac{1}{\sqrt{\varepsilon-\delta}} \Theta(\varepsilon-\delta)
\end{align}
possesses a one-dimensional singularity as $\varepsilon$ approaches the tuning parameter, $\varepsilon \to \delta$.
Brazovskii \cite{Brazovski:75S} argued that the interaction correction due to fluctuations with such a singular density of states changes the second-order mean-field transition into a fluctuation-induced first-order transition.

In the following, we discuss in detail the application of Brazovskii theory to chiral magnets.
We start from the two-particle irreducible (2PI) effective action \cite{Luttinger:60,Baym:61,Cornwall:74}
\begin{align} \label{effaction}
\lefteqn{\Gamma[\vect \phi, \chi] =}
\\\nn&= \mathcal{S}[\vect \phi] + \frac{1}{2} {\rm Tr} \log \chi^{-1} +  \frac{1}{2} {\rm Tr} \left(\chi^{-1}_0 - \chi^{-1} \right) \chi
+ \Gamma_2[\vect \phi, \chi] ,
\end{align}
which is a functional of the field configuration $\vect \phi$ and the propagator $\chi$. The action $\mathcal{S}[\vect \phi] = \mathcal{F}[\vect \phi]/(k_B T)$ is given by the free energy functional of the chiral magnet defined in the previous section.
Here the trace, Tr, should be taken in the functional sense. The bare susceptibility was already defined in Eq.~\eqref{BareSusc}.
Finally, $\Gamma_2$ is the sum of all 2PI vacuum graphs evaluated with the renormalized propagator $\chi$, see Fig.~\ref{fig:Gamma2}.
The effective action has to be minimized such that
\begin{align} \label{EoS}
\frac{\delta \Gamma[\vect \phi, \chi]}{\delta \vect \phi(\vect r)} = 0,\qquad
\frac{\delta \Gamma[\vect \phi, \chi]}{\delta \chi(\vect r,\vect r')} = 0,
\end{align}
which determines the observable field configuration $\vect \phi$ and the susceptibility $\chi$.

\subsection{Hartree-Fock-Brazovskii (HFB) approximation for the effective potential}

\begin{figure}[th]
\includegraphics[width=.5\textwidth]{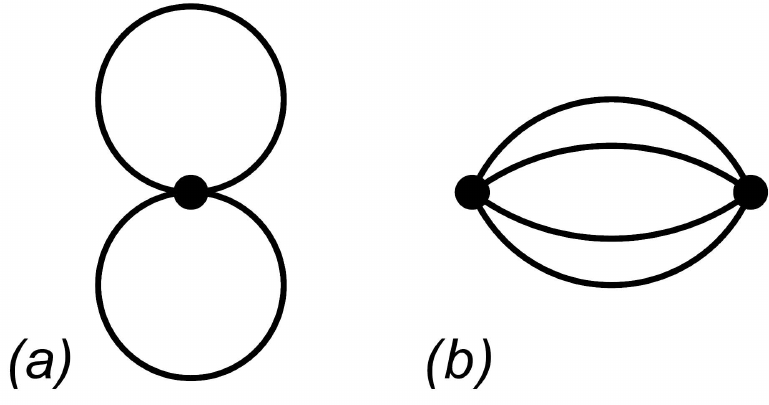}
\caption{Two-particle irreducible diagrams of first (a) and second (b) order in the interaction. The interaction is represented by the dot and the line corresponds to a fluctuation propagator. We approximate $\Gamma_2$ of Eq.~\eqref{effaction} with the lowest order diagram (a).
}
\label{fig:Gamma2}
\end{figure}

The following approximation consists of two steps. First, we will limit ourselves for the 2PI vacuum graphs to the lowest order diagram in Fig.~\ref{fig:Gamma2}(a) corresponding to the Hartree-Fock approximation,
\begin{align} \label{HF}
\Gamma_2 = \frac{u}{4! k_B T} \int d\vect r \left[ \left({\rm tr} \chi(\vect r,\vect r) \right)^2 + 2 {\rm tr} \chi^2(\vect r,\vect r) \right].
\end{align}
The second approximation concerns the form of the susceptibility matrix. It was argued by Brazovskii \cite{Brazovski:75S} that close to the transition the most singular contribution of the fluctuation correction is atrributed to the part of the fluctuation propagator that is diagonal in momentum. We thus make the following Ansatz for the susceptibility
\begin{align} \label{SuscAnsatz0}
\chi^{-1}_{ij}(\vect k, \vect k') &= \chi^{-1}_{ij}(\vect k) \delta_{0, \vect k+\vect k'} + \chi^{-1}_{0, ij}(\vect k, \vect k') (1 - \delta_{0, \vect k+\vect k'})
\end{align}
While the off-diagonal part coincides with the one of the bare susceptibility \eqref{BareSusc2}, we parametrize the diagonal part as
\begin{align} \label{SuscAnsatz}
\chi^{-1}_{ij}(\vect k) = \frac{1}{k_B T}\Big[(J Q^2 + J k^2) \delta_{ij} - i 2 D \varepsilon_{ijn} k_n
\\\nn
+ \Delta_\perp \left(\delta_{ij} - \hat n_i \hat n_j \right)
+ \Delta_\parallel  \hat n_i \hat n_j \Big]
\end{align}
with the two variational parameters $\Delta_\perp$ and $\Delta_\parallel$, and the unit vector $\hat n$. For our purposes it is sufficient to assume that for $\Delta_\perp \neq \Delta_\parallel$ there is only a single preferred orientation given by $\hat n$. At any finite field $H$ in the paramagnetic phase $|\Psi_{\rm hel}|=0$ or in the helimagnetically ordered phase, $|\Psi_{\rm hel}| >0$, for fields $H>H_{c1}$, the unit vector $\hat n$ can be identified with the magnetic field orientation, $\hat n = \hat H$. For $|\Psi_{\rm hel}| >0$ and $H=0$, on the other hand, the orientation is determined by the pitch orientation of the helimagnetic domain so that $\hat n = \hat Q$. However, for $H \approx H_{c1}$ within the ordered phase where cubic anisotropies compete with the magnetic energy to orient the pitch, see Eq.~\eqref{Potcub}, two orientations might be present and the parametrization of Eq.~\eqref{SuscAnsatz} might be insufficient.

Confining ourselves to the critical regime and anticipating that the crossover close to criticality associated with the energy scale $\varepsilon_{\rm cub}$ of Eq.~\eqref{EnergyCub} is basically preempted by the first order transition, we neglect in the following the corrections from cubic anisotropies to Eq.~\eqref{SuscAnsatz}. Furthermore, we apply the Brazovskii approximation  \cite{Brazovski:75S}
\begin{align} \label{BrazAppr}
\chi^{-1}_{0 ij}(\vect k,\vect k') &\approx \chi^{-1}_{0 ij}(\vect k) \delta_{0,\vect k + \vect k'},\quad
\chi^{-1}_{ij}(\vect k,\vect k') \approx \chi^{-1}_{ij}(\vect k) \delta_{0,\vect k + \vect k'} .
\end{align}
in the evaluation of the terms in \eqref{effaction}, i.e., we neglect the components off-diagonal in momentum space which are present in the ordered phase. Eq.~\eqref{BrazAppr} together with Eq.~\eqref{HF} yields the self-consistent Hartree-Fock-Brazovskii (HFB) approximation for the effective potential.

For the evaluation of Eq.~\eqref{effaction}, it is convenient to define the function
\begin{align} \label{DFunction}
D(\Delta_\perp,\Delta_\parallel)  = k_B T \int \frac{d \vect k}{(2\pi)^3} \log \chi^{-1}(\vect k),
\end{align}
with the susceptibility of Eq.~\eqref{SuscAnsatz}. In the following, we will also need derivatives of this function which we denote by a corresponding subscript, e.g., $D_\perp(\Delta_\perp,\Delta_\parallel) =  \partial_{\Delta_\perp} D(\Delta_\perp,\Delta_\parallel)$.
The function $D$ can be separated into a part containing the leading singularity and a part, that is subleading  close to criticality, $D= D_{\rm sing} + D_{\rm sub}$. The singular part is given by
\begin{align}
D_{\rm sing}(\Delta_\perp,\Delta_\parallel) =
\frac{Q^3 k_B T}{\sqrt{2} \pi} \sqrt{\frac{\Delta_\perp + \Delta_\parallel}{J Q^2}} \mathcal{Y}\Big(\frac{\Delta_\perp - \Delta_\parallel}{\Delta_\perp + \Delta_\parallel}\Big)
\end{align}
where the auxiliary function $\mathcal{Y}$ reads explicitly
\begin{align}
\mathcal{Y}(\alpha) = \sqrt{1+\alpha} - \frac{1}{\pi} \int_{-\infty}^{\infty} ds \left(1 - \sqrt{\frac{1+s^2}{\alpha}} \arctan \sqrt{\frac{\alpha}{1+s^2}}\right)
\end{align}
and obeys $\mathcal{Y}(0) = 1$ and $\mathcal{Y}'(0) = \frac{1}{6}$. The subleading part has the property that
its first derivatives have a well-defined limit as $\Delta_{\perp,\parallel} \to 0$.

Using the mean-field Ansatz \eqref{MFAnsatz} for the field configuration, the effective action $\Gamma$ then reduces to the effective potential $\Gamma = V \mathcal{V}_{\rm eff}/(k_B T)$ with
\begin{gather}  \label{EffPotHFB}
\mathcal{V}_{\rm eff}(\Delta_\perp,\Delta_\parallel, \phi_0, \psi_{\rm hel},\psi^*_{\rm hel}) =
\mathcal{V}_0(\phi_0,\psi_{\rm hel},\psi^*_{\rm hel})
+ \frac{1}{2} D(\Delta_\perp,\Delta_\parallel)
\nn\\\label{effPot1}
+ \frac{1}{2} \left(\delta + \frac{u}{3!} (\phi_0^2 + 4 |\psi_{\rm hel}|^2) - \Delta_\perp \right) D _\perp
\\\nn+ \frac{1}{2} \left(\delta + \frac{u}{3!} (3 \phi_0^2 + 2 |\psi_{\rm hel}|^2) - \Delta_\parallel \right) D_\parallel
\\\nn
+ \frac{u}{4!} \Big(
(D _\perp +D _\parallel )^2 + 2 \Big(\frac{1}{2} D ^2_\perp +D ^2_\parallel \Big)
\Big)
\end{gather}
where $\mathcal{V}_0$ is the mean-field potential given in Eq.~\eqref{PotMF}.
We assume that the part of the mean-field potential $\mathcal{V}_{\rm cub}$ of Eq.~\eqref{Potcub} determining the orientation of the pitch $\hat Q$ and the orientation of the homogeneous field $\hat \phi_0$ is already minimized and that in the ordered phase it yields $\hat \phi_0 = \hat Q$ for finite $\phi_0$, which might exclude the regime close to the pitch-flop transition that we do not consider in the following for simplicity.

The effective potential $\mathcal{V}_{\rm eff}$ now is to be minimized with respect to $\Delta_\perp$ and $\Delta_\parallel$, $\phi_0$ and the complex amplitude $\psi_{\rm hel}$. Minimization with respect to the susceptibility parameters, $\partial_{\Delta_\perp}\mathcal{V}_{\rm eff}=0$ and  $\partial_{\Delta_\parallel}\mathcal{V}_{\rm eff}=0$ yield the two equations
\begin{align} \label{Dperp}
\Delta_\perp &= \delta + \frac{u}{3!} (\phi_0^2 + 4 |\psi_{\rm hel}|^2)
+ \frac{u}{3!} \Big(2 D_\perp+  D_\parallel \Big)
\\\label{Dpara}
\Delta_\parallel &= \delta + \frac{u}{3!} (3 \phi_0^2 + 2 |\psi_{\rm hel}|^2)
+ \frac{u}{3!} \Big(D_\perp +  3 D_\parallel\Big)
\end{align}
Minimization with respect to the amplitudes, $\partial_{\phi_0}\mathcal{V}_{\rm eff}=0$ and $\partial_{\psi^*_{\rm hel}}\mathcal{V}_{\rm eff}=0$ gives the equation of states
\begin{align} \label{EoSPhi0}
\Big(r + \frac{u}{3!} (\phi_0^2 + 2 |\psi_{\rm hel}|^2 +  D_\perp +  3 D_\parallel)\Big) \phi_0 &= \gamma H
\\\label{EoSPsi}
\Big(\delta + \frac{u}{3!}(\phi_0^2 + 2 |\psi_{\rm hel}|^2 + 2   D _\perp
+  D _\parallel )\Big) \psi_{\rm hel} &= 0.
\end{align}
Taking the effective potential $\mathcal{V}_{\rm eff}$ at its minimum yields the free energy density in the HFB approximation from which thermodynamic quantities can be evaluated.
A numerical calculation of the specific heat is shown in Fig.~\ref{fig:Fig5} in the {\it main text}.
One finds that the singular fluctuation corrections renormalize the effective potential strongly so that the second-order mean-field transition is converted into a fluctuation-induced first-order transition.

\subsection{Magnetic susceptibility at zero field H=0}
With the help of the susceptibility tensor \eqref{SuscAnsatz0} we can derive the magnetic susceptibility, $\chi_{{\rm mag},ij} = \frac{\gamma^2 \mu_0}{k_B T} \chi_{ij}(0,0)$, following the derivation of the previous section.
As before, we limit ourselves to the magnetic susceptibility in zero field, $H=0$. In the paramagnetic phase, $T > T_c$,
the magnetic susceptibility is isotropic $\chi_{{\rm mag},ij} = \delta_{ij} \gamma^2 \mu_0/(J Q^2 + \Delta)$,
with $\Delta \equiv \Delta_\parallel=\Delta_\perp$. In the ordered phase, $T < T_c$,
we obtain for the magnetic susceptibility of a single domain
\begin{align}
\lefteqn{\chi_{{\rm mag},ij} =}
\\\nn &\frac{\gamma^2 \mu_0}{J Q^2} \Big(\frac{\hat Q_i \hat Q_j }{1+\Delta_{\parallel}/(J Q^2)}
+ \frac{1 + \Delta_\perp/(JQ^2)}{1+ 2 \Delta_\perp/(J Q^2)} (\delta_{ij} - \hat Q_i \hat Q_j )\Big),
\end{align}
where we used that $\Delta_\perp = (u/3) |\psi_{\rm hel}|^2$ which follows from combining the equation of state \eqref{EoSPsi} and Eq.\eqref{Dperp}. Averaging over domains as before, we thus obtain for the longitudinal susceptibility in the case of zero field cooling (ZFC)
\begin{align} \label{SuscZFCBraz}
\lefteqn{\langle \chi_{{\rm mag}} \rangle|_{\rm ZFC} = }
\\\nn&
\frac{\gamma^2 \mu_0}{\varepsilon_{\rm DM}}
\left\{
\begin{array}{lll}
1/(1 + \Delta/\varepsilon_{\rm DM}) & {\rm for} & T > T_c
\\[.7em]
\displaystyle
\frac{1}{3}\Big(\frac{1}{1+\Delta_{\parallel}/\varepsilon_{\rm DM}}
+ 2 \frac{1 + \Delta_\perp/\varepsilon_{\rm DM}}{1+ 2 \Delta_\perp/\varepsilon_{\rm DM}} \Big)
& {\rm for} & T < T_c
\end{array}
\right.
\end{align}
where $\varepsilon_{\rm DM} = J Q^2$, and $\Delta_\perp, \Delta_\parallel$ are functions of the tuning parameter $\delta$ that have to be determined by minimizing the effective potential. Far-away from the transition the mean-field behavior is recovered but close to the transition the Brazovskii renormalization leads to qualitative modifications as discussed in the {\it main text}.
Fig.~\ref{fig:Fig4}(d) in the {\it main text} shows a comparison with the experimental data on MnSi.

\subsection{Brazovskii scaling limit in the critical regime}

In the presence of fluctuation corrections an additional energy scale $\varepsilon_{\rm Gi} \sim (u k_B T Q^2/\sqrt{J})^{2/3}$ emerges. As the transition is approached from high temperatures at $H=0$, the fluctuation corrections become non-perturbative if the tuning parameter $\delta$ reaches the order of $\delta \sim \varepsilon_{\rm Gi}$, which corresponds to the Ginzburg criterion for chiral magnets (provided that $\varepsilon_{\rm Gi} < \varepsilon_{\rm DM}$).

It turns out that in the critical regime the effective potential acquires a scaling form if the Brazovskii energy is much smaller than the DM energy density, $\varepsilon_{\rm Gi} \ll \varepsilon_{\rm DM}$.
This scaling limit is obtained when the function $D$ is approximated by its singular part $D_{\rm sing}$ only. For example, in the paramagnetic phase, $|\psi_{\rm hel}| = 0$, at zero field, $H=\phi_0 = 0$, the equations of state are automatically satisfied, and Eqs.~\eqref{Dperp} and \eqref{Dpara} reduce to a single equation as $\Delta_\perp = \Delta_\parallel \equiv \Delta$ that has a simple form in the scaling limit given by
\begin{align}  \label{BrazEqu}
\Delta &= \delta
+ \frac{\varepsilon_{\rm Gi}^{3/2}}{\sqrt{\Delta}} .
\end{align}
This equation has the Brazovskii-form; the $1/\sqrt{\Delta}$ dependence of the fluctuation correction reflects the singular density of states \eqref{DOS1d}. Eq.~\eqref{BrazEqu} defines the Ginzburg energy density including prefactors to be
\begin{align}  \label{GiScale}
\varepsilon_{\rm Gi} =  \left(\frac{5}{36\pi} \frac{u k_B T Q^3}{\sqrt{JQ^2}}\right)^{2/3}.
\end{align}
With the identification $\Delta = J \kappa^2$, $\delta = J \kappa^2_{\rm MF}$ and $\varepsilon_{\rm Gi} = J \kappa_{\rm Gi}^2$, Eq.~\eqref{BrazEqu} reproduces Eq.~\eqref{eq:kappa_bra} in the {\it main text}.

\subsection{Corrections to the Hartree-Fock-Brazovskii approximation}

The HFB approximation consists of (i) approximating the 2PI vacuum graphs only by the lowest order one, Eq.~\eqref{HF}, and (ii) neglecting the off-diagonal components of the susceptibility matrix, Eq.~\eqref{BrazAppr}. In the following, we discuss the validity of these two approximations separately.

The correction to the Hartree-Fock approximation of $\Gamma_2$ corresponds to the next-to-leading diagram shown in Fig.~\ref{fig:Gamma2}(b). In the scaling limit, we find that this diagram results in a singular correction to the right hand side of Eq.~\eqref{BrazEqu} of the order
\begin{align}
\delta\Delta \sim \frac{\varepsilon_{\rm Gi}^{3}}{\Delta^{3/2} \sqrt{\varepsilon_{\rm DM}}}.
\end{align}
In order to estimate the regime where this correction is negligible we follow Ref.~\cite{Brazovski:75S} and demand that $\delta \Delta/\Delta \ll 1$. Close to the fluctuation-induced first-order transition at $H=0$, the parameters are determined by the Ginzburg scale so that $\Delta \sim \varepsilon_{\rm Gi}$ and this condition becomes
\begin{align}
\frac{\delta \Delta}{\Delta} \sim \sqrt{\frac{\varepsilon_{\rm Gi}}{\varepsilon_{\rm DM}}} = \frac{\kappa_{\rm Gi}}{\kappa_{\rm DM}} \ll 1
\end{align}
The approximation (i) is thus self-consistent as long as the Ginzburg scale is much smaller than the DM energy scale.
In the case of MnSi, we find $\kappa_{\rm Gi}/\kappa_{\rm DM} = \xi_{\rm DM}/\xi_{\rm Gi} \approx 0.5$, so that quantitative corrections due to Fig.~\ref{fig:Gamma2}(b) can be sizeable.

The susceptibility matrix, Eq.~\eqref{BrazAppr}, is non-diagonal in momentum space as the helical order parameter carries momentum. The approximation (ii) neglects the off-diagonal components of the susceptibility within the ordered phase.
This approximation will account for the quasi-1d renormalization arising from the Brazovskii-spectrum close to criticality but it neglects corrections arising from the Goldstone mode within the helimagnetically ordered phase. These latter corrections can become logarithmically large in the isotropic limit if $H=0$ and cubic anisotropies are absent. The dispersion of this Goldstone mode is then anomalously soft \cite{Belitz:06,Radzihovsky:11} resulting in the absence of true long-range order as a consequence of the so-called Landau-Peierls instability familiar from smectic liquid crystals \cite{ChaikinBookS}.
Importantly, the cubic anisotropies in the present case will regularize the logarithmic singularities so that true long-range order sets in at a finite critical temperature. We assume that the cubic anisotropies are small enough, $\varepsilon_{\rm cub} \ll \varepsilon_{\rm Gi}$, so that they can be neglected in the Brazovskii renormalization of the effective potential but are sufficiently large in regularizing the logarithmic singularities within the ordered phase. The study of the residual logarithmic terms and the modifications of thermodynamic anomalies in the limit of vanishing cubic anisotropies will be addressed in a separate publication.

\begin{table}[th!]
\caption{Estimate of microscopic parameters of the Ginzburg-Landau theory of Eqs.~\eqref{model} and \eqref{modelCubAn} close to the critical temperature, see text}\label{tab:1}
\begin{tabular}{@{\vrule height 10.5pt depth4pt  width0pt}p{0.07\textwidth}p{0.05\textwidth}p{0.1\textwidth}}
\hline
$Q$ & $\approx$ & 0.039/\AA
\\
$J$ & $\approx$ & 2.8 meV/\AA
\\
$|J_{\rm cub}|$ & $\approx$ & 0.13 meV/\AA
\\
$u$ & $\approx$ & 0.32 meV/\AA$^3$
\\\hline
\end{tabular}
\end{table}

\begin{table}[th!]
\caption{Estimate of characteristic energy densities, see text.
}\label{table}

\begin{tabular}{@{\vrule height 10.5pt depth4pt  width0pt}p{0.07\textwidth}p{0.1\textwidth}p{0.05\textwidth}}
\hline
units of & $10^{-3} meV$/\AA$^3$
&$\varepsilon_{\rm DM} $
\\\hline
$\varepsilon_{\rm DM}$& $4.3$ & $1$
\\
$\varepsilon_{\rm Gi}$ & $1.0$ & $0.23$
\\
$\varepsilon_{\rm cub}$ & $0.1$ & $0.023$
\\\hline
\end{tabular}
\end{table}

\section{Estimate of parameters and characteristic scales}

We discuss in the following the interpretation of thermodynamics of MnSi close to the critical temperature in terms of the HFB approximation for chiral magnets. In particular, we discuss the values of the microscopic parameters listed in Table \ref{tab:1} of the Ginzburg-Landau theory of Eqs.~\eqref{model} and \eqref{modelCubAn} that have been used for the comparison.

The length of the pitch $Q\approx 0.039/$\AA\, close to the transition temperature was determined with the help of our SANS data,
see Fig.~\ref{fig:Fig4}(b) in the {\it main text}. From the fit of the magnetic susceptibility to Eq.~\eqref{eq:chi} of the {\it main text} we obtained $\chi_0 \approx 0.27$ that can be identified with $\chi_0 = \gamma^2 \mu_0/(J Q^2)$, from which follows for the stiffness $J \approx 2.8 meV$/\AA. This determines the DM energy density $\varepsilon_{\rm DM} \approx 4.3 \times 10^{-3} meV/$\AA$^3$.
From the fit to the magnetic intensity we extracted the cubic anisotropy parameter $\alpha_{\rm cub}^2 = |J_{\rm cub}|/(2J)$. It assumes a maximal value $\alpha_{\rm cub}^2 \approx 0.023$ at the transition from which we can estimate $|J_{\rm cub}| \approx 0.13 meV$/\AA. For the cubic energy scale we obtain $\varepsilon_{\rm cub} = J \alpha_{\rm cub}^2 Q^2 \approx 0.1\times 10^{-3} meV/$\AA$^3$.
The value for the interaction $u$ is more difficult to determine. One possible way is to use the magnetic equation of state of Eq.\eqref{mEoS}. Comparing with magnetization measurements \cite{Bauer:10S} on MnSi at small fields along $\langle 100 \rangle$ at a temperature $T=32$ K we obtain for the interaction constant $u_{\rm Arrot} \approx 0.13 meV$/\AA$^3$. However, experimentally the magnetic equation of state has not the simple mean-field form \eqref{mEoS} probably due to strong Brazovskii renormalizations and the error for the extracted value of $u$ is probably large. For this reason, we used instead the value that follows from the Brazovskii fit to the correlation length that yielded a Ginzburg length $\kappa_{\rm Gi} \approx 0.019/$\AA. This translates to a Ginzburg energy density $\varepsilon_{\rm Gi} = J \kappa_{\rm Gi}^2 \approx 1.0 \times 10^{-3} meV/$\AA$^3$ and, in turn via Eq.~\eqref{GiScale}, to the value for the interaction $u \approx 0.32 meV/$\AA$^3$.

Finally, the temperature dependence of the tuning parameter was extracted via the Curie-Weiss limit of the magnetic susceptibility. Experimentally, one finds sufficiently far above the critical temperature that the dimensionless magnetic susceptibility obeys Curie-Weiss behavior $\chi_{\rm mag} = T_\chi/(T-T_{\rm C})$ with a Curie temperature
\begin{align} \label{Temps}
T_{\rm C} \approx 28.2K\quad {\rm and} \quad T_\chi \approx 0.62 K.
\end{align}
The values were extracted from the asymptotic behavior of the fit to the susceptibility in Fig.~\eqref{fig:Fig5}(a) in the {\it main text}. The temperature dependence of the tuning parameter
$\delta \propto T-T_{\rm MF}$
or, equivalently, $r$ of Eq.~\eqref{model} was determined by adjusting the susceptibility so that the observed Curie-Weiss behavior is recovered. This fixed all microscopic parameters and the resulting HFB approximation
yielded the thermodynamics displayed in Fig.~\eqref{fig:Fig5} in the {\it main text}.

\end{bibunit}

\end{document}